\documentclass[12pt]{JHEP3}

\usepackage{cite}
\usepackage{dsfont}
\usepackage{amsfonts}
\usepackage{amssymb,latexsym}
\usepackage{amsmath,amsbsy}
\usepackage{epsfig,bm}
\usepackage{graphicx,comment}
\usepackage[active]{srcltx}

\usepackage{yfonts}

\def\cA{{\cal A}} \def\cB{{\cal B}}  
  \def\cG{{\cal G}}

\def\Tr{\mathop{\rm Tr}}

\newcommand{\tmfloatcontents}{}
\newlength{\tmfloatwidth}
\newcommand{\tmfloat}[5]{
  \renewcommand{\tmfloatcontents}{#4}
  \setlength{\tmfloatwidth}{\widthof{\tmfloatcontents}+1in}
  \ifthenelse{\equal{#2}{small}}
    {\ifthenelse{\lengthtest{\tmfloatwidth > \linewidth}}
      {\setlength{\tmfloatwidth}{\linewidth}}{}}
    {\setlength{\tmfloatwidth}{\linewidth}}  \begin{minipage}[#1]{\tmfloatwidth}
    \begin{center}
      \tmfloatcontents
      \captionof{#3}{#5}
    \end{center}
  \end{minipage}}

\newcommand{\be}{\begin{equation}}
\newcommand{\ee}{\end{equation}}
\newcommand{\bea}{\begin{eqnarray}}
\newcommand{\eea}{\end{eqnarray}}
\newcommand{\bem}{\begin{multline}}
\newcommand{\eem}{\end{multline}}
\newcommand{\beg}{\begin{gather}}
\newcommand{\eeg}{\end{gather}}

\def\eq#1{{Eq.~(\ref{#1})}}
\def\fig#1{{Fig.~\ref{#1}}}
\newcommand{\ben}{\begin{eqnarray*}}
\newcommand{\een}{\end{eqnarray*}}

\def\peq#1{{(\ref{#1})}}

\graphicspath{{Figs/}}

\newcommand{\tgh}{\textgoth{h}}

\newcommand{\tgt}{\textgoth{t}}

\title{Gravity Dual Corrections to the Heavy Quark Potential at Finite-Temperature}

\author{Hovhannes R. Grigoryan, Yuri V. Kovchegov \\
\vspace{0.1in}

Department of Physics, The Ohio State University, Columbus,
OH 43210, USA \\~~\\

E-mail addresses: \email{grigoryan@physics.osu.edu},
\email{kovchegov.1@asc.ohio-state.edu}

\vspace{0.1in}
}

\abstract{ We apply gauge/gravity duality to compute $1/N^2_c$
  corrections to the heavy quark potentials of a quark--anti-quark pair
  ($Q\bar Q$) and of a quark--quark pair ($QQ$) immersed into the
  strongly coupled ${\cal N} = 4$ SYM plasma.  On the gravity side
  these corrections come from the exchanges of supergravity modes
  between two string worldsheets stretching from the UV boundary of
  AdS space to the black hole horizon in the bulk and smeared over $S^5$.
  We find that the contributions to the $Q\bar Q$ potential coming
  from the exchanges of all of the relevant modes (such as dilaton,
  massive scalar, 2-form field, and graviton) are all attractive,
  leading to an attractive net $Q\bar Q$ potential.  We show that at
  large separations $r$ and/or high-temperature $T$ the potential is
  of Yukawa-type, dominated by the graviton exchange, in agreement
  with earlier findings.  On the other hand, at small-$r \, T$ the
  $Q\bar Q$ potential scales as $\sim (1/r) \ln (1/r\, T)$.  In the
  case of $QQ$ potential the 2-form contribution changes sign and
  becomes repulsive: however, the net $QQ$ potential remains
  attractive. At large-$r \, T$ it is dominated by the graviton
  exchange, while at small-$r\, T$ the $QQ$ potential becomes
  Coulomb-like.
}


\preprint{\today}

\begin{document}

%%%%%%%%%%%%%%%%%%%%%%%%%%%%%%%%%%%%%%%%%%%%%%%%%%%%%%%%%%%%%%%%%%%%%%%%%%%%%%%
%%%%%%%%%%%%%%%%%%%%%%%%%%%%%%%%%%%%%%%%%%%%%%%%%%%%%%%%%%%%%%%%%%%%%%%%%%%%%%%

\section{Introduction}
\label{intro}

Heavy quark potential in the vacuum of the ${\cal N} =4$
supersymmetric Yang-Mills (SYM) theory
\cite{Maldacena:1998im,Rey:1998ik} was one of the first results
obtained using the Anti-de Sitter space/Conformal Field Theory
(AdS/CFT) correspondence
\cite{Maldacena:1997re,Gubser:1998bc,Witten:1998qj,Witten:1998zw,Aharony:1999ti}.
The first attempt to generalize the calculation
\cite{Maldacena:1998im,Rey:1998ik} to the case of the ${\cal N}=4$ SYM
medium at finite temperature ($T$) was made in
\cite{Rey:1998bq,Brandhuber:1998bs} shortly thereafter.  The authors
of \cite{Rey:1998bq,Brandhuber:1998bs} identified two string
configurations contributing to the finite-T heavy quark potential in
the AdS/CFT framework, shown below in \fig{strings}, and calculated
their contributions. Defining the heavy quark potential as the free
energy of the quark--anti-quark system immersed in the thermal bath,
and assuming that the two string configuration from \fig{strings}
contribute to the free energy on equal footing, the authors of
\cite{Rey:1998bq,Brandhuber:1998bs} obtained a potential which had a
kink when plotted as a function of the quark--anti-quark separation
$r$. The kink (a derivative discontinuity) was due to a transition
from the regime when one string configuration dominated the potential
to the regime where another one was more important.

The kink feature of the result obtained in
\cite{Rey:1998bq,Brandhuber:1998bs} appears to be an artifact of the
large-$N_c$ approximation, as was noted in
\cite{Gross:1998gk,Bak:2007fk}. In \cite{Gross:1998gk,Bak:2007fk} it
was argued that there are of the order of $N_c^2$ configurations with
two straight strings (shown in the right panel of \fig{strings}), such
that the two configurations in \fig{strings} do not come in on equal
footing, contrary to the assumption made in
\cite{Rey:1998bq,Brandhuber:1998bs}. Here $N_c$ is the number of D3
branes, and $N_c^2$ configurations may be interpreted as resulting
from Chan-Paton indices of the two strings, each string having $N_c$
of indices (see also \cite{Aharony:1998qu}, where $N_c^2$
configurations emerge from the topological $Z_N \times Z_N$ symmetry
of thermal SYM theory). This large number of string configurations
enhances the terms which would be otherwise subleading.  As was
pointed out in \cite{Bak:2007fk}, while the exchanges of supergravity
fields between the straight strings shown in \fig{exchange} are
normally $N_c^2$-suppressed, they become leading-order due to the fact
that there are $N_c^2$ string configurations to enhance their
contributions.  The primary goal of \cite{Bak:2007fk} was
determination of Debye mass, defined as the screening mass for
chromo-electric modes in the plasma following \cite{Arnold:1995bh}:
therefore, for the heavy quark potential, the authors of
\cite{Bak:2007fk} used the glueball masses calculated for QCD$_3$ from
AdS/CFT correspondence in \cite{Csaki:1998qr,Brower:2000rp}, to obtain
only the exponential form of the large-$r$ asymptotics of the
contribution coming from the supergravity field exchanges to the heavy
quark potential.

The question of the definition of the heavy quark potential at
finite-$T$ in the context of AdS/CFT was raised in
\cite{Albacete:2008dz}. There it was pointed out that one can define
{\sl two} heavy quark potentials: color-singlet and color-adjoint
\cite{McLerran:1981pb,Nadkarni:1986cz}. It was further conjectured in
\cite{Albacete:2008dz} that the hanging string configuration (left
panel in \fig{strings}) may correspond to the color-singlet potential,
while the two-strings configuration (right panel in \fig{strings}) may
give the adjoint potential (which is zero in the large-$N_c$ limit).
It was further suggested in \cite{Albacete:2008dz} that the singlet
potential at large-$r\, T$, for which the solution corresponding to
the hanging string configuration becomes complex, may be obtained by
analytically continuing the solution into complex domain. Such
analytic continuation led to an absorptive potential which had both
real and imaginary parts, with the real part falling off as a power of
$r$ at very large-$r$, in some similarity to the weak-coupling
perturbative results
\cite{Gale:1987en,Laine:2006ns,Brambilla:2008cx,Burnier:2009bk,Baier:1994et,Peigne:1994dn,Taliotis:2010kx}.
The issues associated with different definitions of the heavy quark
potential are presented below in Sec. \ref{two}, where we present both
definition of the finite-$T$ heavy quark potential, and discuss the
pros and cons of using each of them.

The central goal of this paper is to find the full contribution to the
heavy quark potential coming from the exchanges of supergravity fields
between two straight strings oriented in opposite directions, as shown
in \fig{exchange} below. In the language of \cite{Albacete:2008dz}
this contribution would correspond to the color-adjoint potential. The
calculation is presented in Sec.~\ref{QQbar}, and is carried out in
Euclidean time, making it free of subtleties of the real-time
formalism of thermal field theories.  We concentrate on the case of
the strings smeared over the $S^5$, which correspond to Polyakov lines
defined below in \eq{L4} averaged over the directions of their
coupling to six scalars of ${\cal N}=4$ SYM.  In this case the
supergravity exchanges are limited to $k=0$ Kaluza-Klein (KK) modes.
The contributions of all the relevant supergravity modes (dilaton,
massive scalar, 2-form field and graviton) are calculated in detail in
Sec. \ref{QQbar}. We find that all the exchanges give attractive
contributions to the $Q\bar Q$ potential.  The net result for the
$Q\bar Q$ potential in momentum space is shown in \fig{totalpot}
(lower line). The potential turns out to be attractive at all values
of $r \, T$ (which is the only relevant dimensionless parameter in the
problem). The large-$r \, T$ asymptotics of the net $Q\bar Q$
potential is given by the exponential decay of \eq{net_large_r}, with
the exponent determined by the first pole of the momentum ($q$)-space
potential along the imaginary-$q$ axis (in agreement with
\cite{Bak:2007fk}). The pole corresponds to the glueball mass in
QCD$_3$ calculated in \cite{Csaki:1998qr,Brower:2000rp}.  We also find
the residue of the pole (the factor in front of the exponential). The
small-$r \, T$ asymptotics of the potential is rather peculiar, since
it is not quite Coulomb-like, as can be seen from \eq{net-small-r}
below: the Coulomb $1/r$ term is multiplied by $\ln (1/r\, T)$,
generating a potential singularity in the $T \rightarrow 0$ limit.
This singularity might be related to the instability of our non-BPS
string configuration.  We discuss how the quantum effects, such as
string fluctuations, may come in to regulate this logarithmic
divergence at very small-$r\, T$.

Using the developed machinery, we also calculate the quark--quark $QQ$
potential in Sec. \ref{QQ}, for which there is no hanging (U-shaped)
string configuration, and the only contribution comes from the
exchanges shown in \fig{qq_pot} between the two strings oriented in
the same direction. The only difference between the $QQ$ potential and
the adjoint contribution to the $Q\bar Q$ potential is in the sign of
the NS 2-form contribution, which in the $QQ$ case becomes repulsive.
The net $QQ$ potential is still attractive, and is plotted in momentum
space in \fig{totalpot} (upper line). It is interesting to note that
at short distances the $QQ$ potential becomes Coulomb-like, as shown
in \eq{netVqq_T0_r}. The absence of the logarithmic singularity that
we found earlier in the $Q\bar Q$ system can be related to the fact
that the system of two parallel branes is (BPS) stable.
 
We summarize in Sec. \ref{sum} by suggesting that the 2-form exchange,
which changes sign in going from $Q\bar Q$ to $QQ$, may correspond to
the chromo-electric modes in the gauge theory, while the exchanges of
all other supergravity fields (the scalars and the graviton) may be
interpreted as being mainly due to chromo-magnetic modes. Since both
the $Q\bar Q$ and $QQ$ potentials are attractive we infer that at
strong coupling magnetic modes are more important. We conclude by
discussing the possibilities of improving on our calculation,
including possibly making it more QCD-like.

%%%%%%%%%%%%%%%%%%%%%%%%%%%%%%%%%%%%%%%%%%%%%%%%%%%%%%%%%%%%%%%%%%%%%%%%%%%%%%%
%%%%%%%%%%%%%%%%%%%%%%%%%%%%%%%%%%%%%%%%%%%%%%%%%%%%%%%%%%%%%%%%%%%%%%%%%%%%%%%

\section{Two definitions of the heavy quark potential at finite temperature}
\label{two}

Here we will present possible interpretations of the observable we
would like to calculate.

%%%%%%%%%%%%%%%%%%%%%%%%%%%%%%%%%%%%%%%%%%%%%%%%%%%%%%%%%%%%%%%%%%%%%%%%

\subsection{Singlet and octet potentials}

Imagine immersing a very heavy quark and anti-quark into a static
thermal medium of an $SU (N_c)$ gauge theory. (For a nice pedagogical
presentation of the topic, along with some weak-coupling calculations,
we refer the reader to \cite{Nadkarni:1986cz}.) Working in the
Euclidean time one defines the Polyakov loop operator at a spatial
location $\vec r$ by
\begin{align}\label{L}
  L ({\vec r}) \, = \, \text{P} \exp \left( i \, g \,
    \int\limits_0^\beta d \tau \, A_0 ({\vec r}, \tau )\right)
\end{align}
with $\tau$ the Euclidean time, $A_0$ the temporal component of the
gauge field, $\beta = 1/T$, and $T$ the temperature.  Using Fierz
identity one may write
\begin{align}
  \label{Fierz}
  \text{Tr} \, L^\dagger (0) \ \text{Tr} \, L ({\vec r}) \, = \, 2 \,
  \text{Tr} \left[ t^a \, L^\dagger (0) \, t^a \, L ({\vec r}) \right]
  + \frac{1}{N_c} \, \text{Tr} \left[ L^\dagger (0) \, L ({\vec r})
  \right],
\end{align}
where $t^a$ are the generators of $SU (N_c)$ in the fundamental
representation, and $\vec r$ and $0$ are spatial locations of the
quark and the anti-quark. One then defines the {\sl color-singlet}
heavy quark potential by
\begin{align}
  \label{singlet_def}
  e^{- \beta \, V_1 (r)} \, \equiv \, \frac{1}{N_c} \, \left\langle
    \text{Tr} \left[ L^\dagger (0) \, L ({\vec r}) \right]
  \right\rangle_c,
\end{align}
and the {\sl color-adjoint} potential by
\begin{align}
  \label{octet_def}
  e^{- \beta \, V_{adj} (r)} \, \equiv \, \frac{2}{N_c^2 -1} \,
  \left\langle \text{Tr} \left[ t^a \, L^\dagger (0) \, t^a \, L
      ({\vec r}) \right] \right\rangle_c.
\end{align}
Here the angle brackets denote the expectation value of the operators
in the thermal bath and the matrix elements are normalized to one for
the case of no interaction. The distance between the quark and the
anti-quark is $r = |{\vec r}|$.  The free energy of the
quark--anti-quark pair in a thermal bath is given by
\begin{align}
  \label{free_energy}
  e^{- \beta \, F (r)} \, = \, \frac{1}{N_c^2} \, \left\langle
    \text{Tr} \, L^\dagger (0) \ \text{Tr} \, L ({\vec r})
  \right\rangle_c.
\end{align}
The subscript $c$ in the above equations implies that we only keep the
connected part of the correlator. This means that the contributions to
the quark--anti-quark free energy and to the potentials due to
self-interactions of the quarks are subtracted out. (The trivial
disconnected part of the correlator due to the contribution without
any (self-)interactions that gives $1$ in the perturbative expansion
of the correlator is included on both sides of \peq{free_energy}.)

Using the definitions \peq{singlet_def}, \peq{octet_def} and
\peq{free_energy} in \eq{Fierz} we write
\cite{McLerran:1981pb,Nadkarni:1986cz}
\begin{align}\label{decomp}
  e^{- \beta \, F (r)}\, = \, \frac{1}{N_c^2} \, \left[ e^{- \beta \,
      V_1 (r) } + (N_c^2 -1) \, e^{- \beta \, V_{adj} (r) } \right].
\end{align}
We see that the quark--anti-quark free energy consists of the singlet
and adjoint contributions. One can show \cite{Nadkarni:1986cz} that at
weak coupling and in the large-$N_c$ limit the singlet potential $V_1
(r)$ is of the order $g^2 \, N_c$ with $g$ the gauge coupling, that is
\begin{align}
  \label{V1coupl}
  V_1 (r) \bigg|_{\lambda \ll 1, \, N_c \gg 1} \, \sim \, \lambda,
\end{align}
where $\lambda = g^2 \, N_c$ is the 't Hooft coupling. The singlet
potential is also {\sl attractive} at weak coupling
\cite{Nadkarni:1986cz}, and maps onto the standard vacuum Coulomb
potential at $T=0$. The adjoint (octet) potential $V_{adj} (r)$ is
{\sl repulsive} at weak coupling \cite{Nadkarni:1986cz}, and is also
subleading in the large-$N_c$ limit, such that
\begin{align}
  \label{V8coupl}
  V_{adj} (r) \bigg|_{\lambda \ll 1, \, N_c \gg 1} \, \sim \,
  \frac{\lambda}{N_c^2}.
\end{align}
In the non-perturbative case, \eq{singlet_def} is used in lattice QCD
simulations to determine the singlet heavy quark potential at
finite-$T$ (see e.g. \cite{Bazavov:2009us}).

While the separation of the quark--anti-quark free energy into singlet
and adjoint components appears to be possible in perturbative
calculations order-by-order in the coupling
\cite{Petreczky:2005bd,Brambilla:2010xn}, it is less clear how to
accomplish this decomposition in the general non-perturbative case.
Indeed the definitions of $V_1 (r)$ and $V_{adj} (r)$ in
Eqs.~\peq{singlet_def} and \peq{octet_def} are not gauge-invariant.
(The sum of the two contributions, and, therefore, the free energy, is
gauge-invariant, as follows from \eq{free_energy}.) We hence have a
question of whether the singlet and adjoint potential are properly
defined.

%%%%%%%%%%%%%%%%%%%%%%%%%%%%%%%%%%%%%%%%%%%%%%%%%%%%%%%%%%%
\FIGURE{\includegraphics[width=4cm]{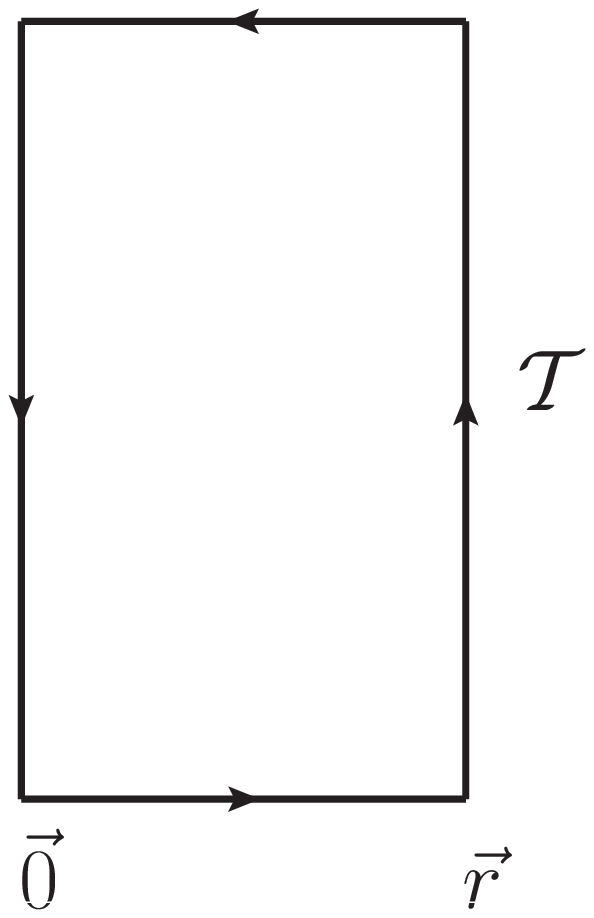}
  \caption{Spatial Wilson loop described in the text.}
  \label{wilson}
}
%%%%%%%%%%%%%%%%%%%%%%%%%%%%%%%%%%%%%%%%%%%%%%%%%%%%%%%%%%%

One possible resolution of this problem is to define the color-singlet
potential using the real (Minkowski) time formulation of thermal field
theory. Define the temporal Wilson loop $W$, which is a rectangle in
the time--$\vec r$ plane, with the spatial sides connecting quark and
anti-quark, and the temporal sides having length $\cal T$, as depicted
in \fig{wilson}. The singlet heavy quark potential can then be defined
by\footnote{Clearly this definition of the singlet potential is
  gauge-invariant, and is, in general, different from the definition
  in \eq{singlet_def}. However, the definition \peq{sing_def2} of $V_1
  (r)$ does agree with that of \eq{singlet_def} both in the
  lowest-order weak-coupling limit and in the strong-coupling AdS/CFT
  calculations.}
\begin{align}
  \label{sing_def2}
  V_1 (r) \, = \, \lim_{{\cal T} \rightarrow \infty} \, \frac{i}{{\cal
      T}} \, \ln W.
\end{align}
After this definition of the singlet potential, and, after defining
the quark--anti-quark free energy (in Euclidean time) with the help of
\eq{free_energy}, one may use \eq{decomp} as the definition of the
adjoint potential $V_{adj} (r)$.  Below, when calculating the adjoint
potential, we will refer to the object defined by \eq{decomp} with the
help of Eqs.~\peq{sing_def2} and \peq{free_energy}.

\begin{figure}[h]
\centering
\includegraphics[width=10cm]{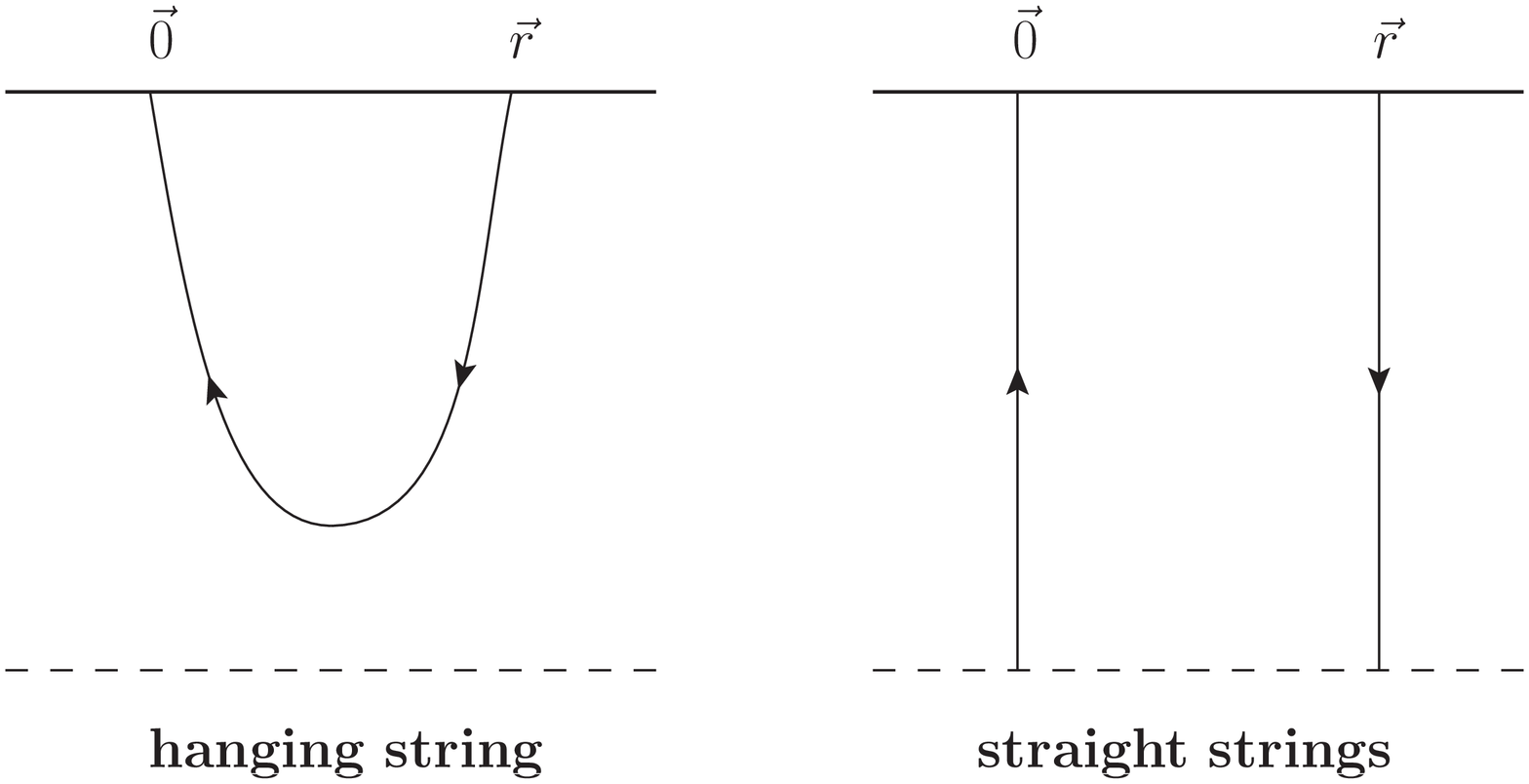}
\caption{Two configurations of open strings contributing to 
  Polyakov loop correlator in an AdS/CFT calculation. Solid horizontal
  line denotes the UV boundary of the AdS space, while the dashed line
  denotes the location of the black hole horizon. The arrows on the
  strings denote their orientations.}
\label{strings}
\end{figure}

If one wants to calculate the Polyakov loop correlator
\peq{free_energy} (with $L$ properly re-defined, as shown in \eq{L4}
below) in the strongly-coupled ${\cal N} =4$ SYM theory using AdS/CFT
correspondence, two distinct string configurations have to be
considered, as shown in \fig{strings}. The first AdS/CFT calculations
of the finite-$T$ heavy quark potential, performed in
\cite{Rey:1998bq,Brandhuber:1998bs} following the zero-temperature
calculation of \cite{Maldacena:1998im,Rey:1998ik}, found the
contribution of the hanging string configuration from the left panel
of \fig{strings}. The leading-order contribution of the right panel in
\fig{strings} is trivial, since, after subtracting the
self-interactions of the quarks, which correspond to the actions of
the two straight strings, we are left with zero answer for the
renormalized action of this configuration, which is what one should
expect in the non-interacting case. As we pointed out above there are
of the order of $N_c^2$ configurations of straight strings.  The setup
appears to be similar to the decomposition of \eq{decomp}: just like
in the gauge theory where one has one color-singlet configuration and
$N_c^2 -1$ color-adjoint configurations, there is one hanging string
configuration and order-$N_c^2$ straight string configurations in
\fig{strings}.  Moreover, the AdS/CFT result for the Polyakov loops
correlator can be written as \cite{Albacete:2008dz}
\begin{align}\label{LL2}
  \frac{1}{N_c^2} \, \left\langle \text{Tr} \, L^\dagger (0) \ 
    \text{Tr} \, L ({\vec r}) \right\rangle_c \propto \frac{e^{-
      S_{hanging}} + (N_c^2-1) \, e^{- S_{straight}}}{N_c^2}
\end{align}
with $S_{hanging}$ and $S_{straight}$ the actions of the hanging and
straight string configurations, with the self-interactions removed by
renormalization. (Indeed we only know that the coefficient in front of
the second term on the right of \eq{LL2} is of the order of $N_c^2$
and we do not have control over $-1$, which we put there only to
normalize the right-hand-side to one in the case of no interactions.)
Comparing Eqs.~\peq{LL2} and \peq{decomp} allowed the authors of
\cite{Albacete:2008dz} to suggest that the action of the hanging
string configuration gives the singlet potential, while the action of
the straight strings configuration gives the adjoint potential.

%%%%%%%%%%%%%%%%%%%%%%%%%%%%%%%%%%%%%%%%%%%%%%%%%%%%%%%%%%%%%%%%%%%%
%%%%%%%%%%%%%%%%%%%%%%%%%%%%%%%%%%%%%%%%%%%%%%%%%%%%%%%%%%%%%%%%%%%

Since, as we have already noted, the decomposition of \eq{decomp} is
not gauge invariant, the suggestion of \cite{Albacete:2008dz} still
needs to be proven. However, let us use our real-time definition of
the singlet potential from \eq{sing_def2}. For a Wilson loop from
\fig{wilson} with a very long temporal extent $\cal T$, the AdS/CFT
calculation, carried out in the Lorentzian-signature metric, would
only contain the hanging string configuration. This is clear since the
straight-strings configuration is impossible in Lorentzian-signature
AdS$_5$: the orientation of the string should be the same throughout
the single string world-sheet of this configuration, while
orientations of the strings connecting to the quark and anti-quark are
opposite, as shown in the right panel of \fig{strings}, making this
configuration impossible in real time. (In addition to that, in case
of Lorentzian-signature black hole the straight-strings world-sheet
would not be simply connected, indicating additional $1/N_c^2$
suppression.)  Since the Nambu--Goto action of the static hanging
string is independent of whether we work in Lorentzian or Euclidean
signature metrics, the potential $V_1 (r)$ defined by \eq{sing_def2}
is going to be the same as found in
\cite{Rey:1998bq,Brandhuber:1998bs}. Thus, using our definition of the
singlet potential \peq{sing_def2}, we see that the hanging string
configuration does indeed give a singlet potential identical to that
in \eq{singlet_def}, yielding \cite{Rey:1998bq,Brandhuber:1998bs}
\begin{align}
  \label{sing_lc}
  V_1 (r) \bigg|_{N_c \gg  \lambda \gg 1} \, \sim \, \sqrt{\lambda}.
\end{align}
The fact that this potential is of the same order in $N_c$ as the
weakly-coupled singlet potential in \eq{V1coupl} indicates consistency
of our conclusions with perturbative calculations.

%%%%%%%%%%%%%%%%%%%%%%%%%%%%%%%%%%%%%%%%%%%%%%%%%%%%%%%%%%%
\FIGURE{\includegraphics[width=5cm]{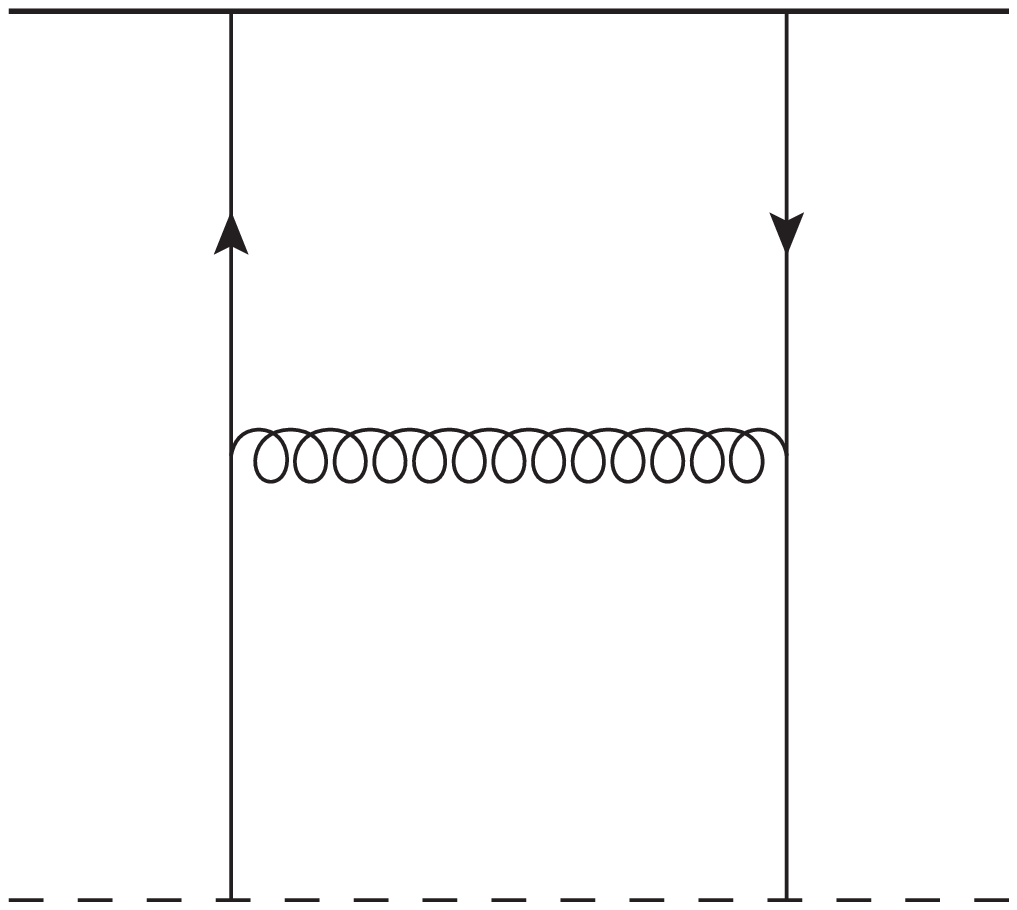}
  \caption{A correction to the straight-strings configuration from \fig{strings} 
    due to exchange of a supergravity field, which, as we will see
    below, can be either a dilaton, massive scalar, 2-form, or a
    graviton, all of which are denoted here by a cork-screw line.}
  \label{exchange}
}
%%%%%%%%%%%%%%%%%%%%%%%%%%%%%%%%%%%%%%%%%%%%%%%%%%%%%%%%%%%

Using \eq{decomp}, and remembering again that the hanging string
configuration gives the singlet potential regardless of whether we
work in real or imaginary time, we see that the straight strings
configuration from \fig{strings} gives the adjoint potential. At the
leading-$N_c^2$ order without any interactions between the strings the
adjoint potential defined this way is zero. As was suggested in
\cite{Bak:2007fk}, a non-zero contribution to $S_{straight}$ comes
from the interactions of the strings due to exchanges of supergravity
fields, as shown in \fig{exchange}. Since the coupling of the
supergravity fields to the string world-sheet is of the order of
$1/N_c$, the contribution to the two-strings action due to the
exchanges shown in \fig{exchange} is order-$1/N_c^2$. Exchanges of
supergravity fields between various string configurations have been
considered before
\cite{Berenstein:1998ij,Danielsson:1998wt,Janik:1999zk}, yielding the
interaction of the order of $\lambda/N_c^2$. Identifying the diagram
in \fig{exchange} as the leading non-trivial contribution to the
adjoint heavy quark potential, we conclude that at strong coupling
\begin{align}
  \label{V8_lc}
  V_{adj} (r) \bigg|_{N_c \gg \lambda \gg 1} \, \sim \,
  \frac{\lambda}{N_c^2}.
\end{align}
Comparing this result to \eq{V8coupl} we observe that, surprisingly,
for the adjoint potential the power of 't Hooft coupling $\lambda$
does not change in going from small to large $\lambda$! This is unlike
the case of the singlet potential in which, just like in many other
AdS/CFT results, the power of the coupling changes from $\lambda$ to
$\sqrt{\lambda}$ as we increase it from very small to very large, as
can be seen from Eqs.~\peq{V1coupl} and \peq{sing_lc}.

In the terminology of the two heavy quark potentials defined above,
the goal of this paper is to calculate the contribution to the adjoint
potential depicted in \fig{strings}.

%%%%%%%%%%%%%%%%%%%%%%%%%%%%%%%%%%%%%%%%%%%%%%%%%%%%%%%%%%%%%%%%%%%%%%%%

\subsection{One unified potential}
\label{One}

There exists an alternative to the singlet and adjoint potential
decomposition.\footnote{We would like to thank Larry Yaffe for a
  discussion on this subject.} One can simply define one unified heavy
quark potential $V (r)$ as the free energy of the quark--anti-quark pair
in the plasma, such that \cite{McLerran:1980pk}
\begin{align}
  \label{one_pot}
  e^{- \beta \, V (r)} \, = \, \frac{1}{N_c^2} \, \left\langle
    \text{Tr} \, L^\dagger (0) \ \text{Tr} \, L ({\vec r})
  \right\rangle_c.
\end{align}
The strength of this definition is that the heavy quark potential
defined this way is manifestly gauge-invariant. 

However, it appears difficult to find an intuitive physical
interpretation for the single unified potential defined in
\eq{one_pot}. To see this, let us start with Eqs.~\peq{free_energy}
and \peq{decomp}, which, together with \eq{one_pot} give
\begin{align}
  \label{one_pot_dec}
  e^{- \beta \, V (r)} \, = \, \frac{1}{N_c^2} \, \left[ e^{- \beta \,
      V_1 (r) } + (N_c^2 -1) \, e^{- \beta \, V_{adj} (r) } \right].
\end{align}
Remembering from the previous Section that $V_1 (r) \sim N_c^0$ and
$V_{adj} (r) \sim 1/N_c^2$ (both at small and large coupling), we
expand the right-hand-side of \eq{one_pot_dec} in powers of $1/N_c^2$
to obtain \cite{Bak:2007fk}
\begin{align}
  \label{one_pot_exp}
  e^{- \beta \, V (r)} \, = \, 1 + \frac{1}{N_c^2} \, \left[ e^{-
      \beta \, V_1 (r) } -1 \right] - \beta \, V_{adj} (r) + O \left(
    \frac{1}{N_c^4} \right).
\end{align}
We see that the unified potential $V (r)$ has to be $N_c^2$-suppressed
and equal to
\begin{align}
  \label{one_pot_exp2}
  V (r) \, = \, \frac{1}{N_c^2} \, \frac{1}{\beta} \, \left[ 1 - e^{-
      \beta \, V_1 (r) } \right] + V_{adj} (r) + O \left(
    \frac{1}{N_c^4} \right).
\end{align}
It seems a bit counter-intuitive that the heavy quark potential at $T
\neq 0$ should be $N_c^2$-suppressed, while the $T=0$ potential is
not. Also, in the $T \rightarrow 0$ limit, the singlet potential $V_1
(r)$ maps smoothly onto the vacuum potential both at small
\cite{Nadkarni:1986cz} and large \cite{Maldacena:1998im,Rey:1998ik}
couplings: it is clear from \eq{one_pot_exp2} that the potential $V
(r)$ instead goes to infinity in the $T \rightarrow 0$ limit.
Moreover, at small quark--anti-quark separations $r$, both in QCD and
in ${\cal N} =4$ SYM theory, the singlet potential is $V_1 (r) \sim
-1/r$ (while $V_{adj} (r) \sim 1/r$ in QCD \cite{Nadkarni:1986cz}, and
$V_{adj} (r) \sim -(1/r) \ln (1/r\, T)$ in ${\cal N} =4$ SYM at
large-$\lambda$, as we will show below). However, as one can infer
from \eq{one_pot_exp2}, the potential $V (r)$ diverges exponentially
at small-$r$ as $- \exp (\text{const} \, \beta/r)$, i.e., it becomes
very strongly attractive: the physical origin of this behavior is not
clear.

Despite the problems with its interpretation, the potential $V (r)$ is
well-defined. As we have already mentioned, in our calculation below
we will find the contribution of the exchanges of supergravity fields
between the two straight strings pictured in \fig{exchange}. In the
language of the unified potential $V (r)$ in \eq{one_pot_exp2}, we
will be constructing one of the contributions to this potential
($V_{adj} (r)$) at order-$1/N_c^2$, while the other contribution
(coming from $V_1 (r)$) was previously found in
\cite{Rey:1998bq,Brandhuber:1998bs}.

%%%%%%%%%%%%%%%%%%%%%%%%%%%%%%%%%%%%%%%%%%%%%%%%%%%%%%%%%%%%%%%%%%%%%%%%%%%%%%%
%%%%%%%%%%%%%%%%%%%%%%%%%%%%%%%%%%%%%%%%%%%%%%%%%%%%%%%%%%%%%%%%%%%%%%%%%%%%%%%

\section{The adjoint contribution to the $Q\bar Q$ potential}
\label{QQbar}

We want to find the contribution to the correlator of two Polyakov
loops
$\left\langle \text{Tr} \, L^\dagger (0) \ \text{Tr} \, L ({\vec r}) \right\rangle_c$ 
in ${\cal N} =4$ SYM theory at strong 't Hooft coupling at
order-$1/N_c^2$.
%-%
On the string theory or gravity side, these contribution will be
coming from the interactions between the two straight strings in \fig{exchange} .
%-%
In ${\cal N} =4$ SYM theory the definition of the Polyakov loop
operator suitable for AdS/CFT calculations is different from \eq{L},
and is given by \cite{Maldacena:1998im,Berenstein:1998ij} (in
Euclidean time)
\begin{align}
  \label{L4}
  L_{{\cal N} =4} ({\vec r}) \, = \, \text{P} \exp \left( i \, g \,
    \int\limits_0^\beta d \tau \, \left[ A_0 ({\vec r}, \tau ) - i \,
      \theta^I ({\vec r}, \tau) \, X^I ({\vec r}, \tau) \right]
  \right),
\end{align}
where $X^I ({\vec r}, \tau)$ with $I = 1, \ldots , 6$, are the six
scalar fields of ${\cal N} =4$ SYM, and $\theta^I$ is a point on the
unit five-sphere, which, for simplicity, will be taken to be
independent of time. To make our results more QCD-like we will try to
reduce their dependence on the scalar fields $X^I$ by averaging
$L_{{\cal N} =4}$ over all $S^5$ angles $\theta^I$.

At low energies type IIB string theory compactified on $AdS_5 \times
S^5$ contains a spectrum of supergravity modes. In the bosonic sector,
the lightest modes are graviton $h_{MN}$, dilaton $\phi$, axion $C_0$,
Neveu--Schwarz (NS) $B_{MN}$ and Ramond--Ramond (RR) $C_{MN}$ 2-form
fields ($M,N = 1, \ldots , 5$), and RR 4-form field $C_4$ with a
self-dual field strength.  The spectrum also contains an infinite
tower of massive scalars, $t^k$ ($k \geq 0$) and $s^k$ ($k \geq 2$)
with $k$ the index of KK modes
\cite{Kim:1985ez,Arutyunov:1999en,Arutyunov:1999fb}.  We are
interested in those modes that couple to a string world-sheet (with
the coupling that is least suppressed in $N_c$). The smearing of the
string over the $S^5$ will effectively eliminate the coupling of the
string world-sheet to modes that carry non-zero KK charge.  In what
follows, we will neglect the contribution of the fermionic sector to
the interaction of the two string world-sheets, since it is
$N_c^2$-suppressed compared to that of the bosonic fields.

We want to find the quadratic fluctuations of those fields that couple
to the string world-sheet as shown in \fig{exchange} in the background
of the AdS Schwarzschild black hole (AdSSBH) metric
\begin{align}\label{AdSSBH}
  ds^2 &= \, g_{MN} \, dx^M \, dx^N \, = \, \frac{1}{z^2}\left[f(z) \,
    d\tau^2 + dx^idx^i + \frac{1}{f(z)} \, dz^2 \right] \ , \ \ \ \ 
  f(z) = 1 - \frac{z^4}{z^4_h} \ ,
\end{align}
where $z \in (0,z_h)$ is the direction along the extra fifth
dimension, $x^i = (x^1,x^2,x^3)$ are spacial coordinates, $\tau$ is
the Euclidean time with period $\beta = \pi z_h$, and $z_h$ determines
the position of the black hole horizon, that is related to temperature
as follows: $T = 1/\beta = 1/(\pi z_h)$.

To find the correlator of Polyakov loops at order-$1/N_c^2$ we will
only need to consider exchanges of the fields that couple to string
world-sheet at the tree level. Such fields are graviton, dilaton,
massive scalars, and NS/RR two forms.  The RR 2-form does not couple
directly to a string world-sheet but it mixes with NS-NS 2-form,
making it effectively massive \cite{Kim:1985ez,Brower:2000rp}.  Notice
that at the tree-level the axion only couples to the world-sheet
fermions, which leads to a higher suppression in $N_c$.  Since we are
interested in order-$1/N_c^2$ corrections, we will ignore the
contribution to the correlator coming from the exchange of the axion.

As we argued above, the quantity we would like to calculate is
\begin{align}
  \label{goal}
  \frac{\left\langle \text{Tr} \, L^\dagger (0, {\vec \theta}') \ 
      \text{Tr} \, L ({\vec r}, {\vec \theta})
    \right\rangle}{\left\langle \text{Tr} \, L^\dagger (0, {\vec
        \theta}') \right\rangle \, \left\langle \text{Tr} \, L ({\vec
        r}, {\vec \theta}) \right\rangle} \Bigg|_{\text{adjoint}} \, =
  \, e^{-S_{straight}} \, = \, 1 - S_{straight} + O \left(
    \frac{1}{N_c^4} \right),
\end{align}
where we assume that the quark and anti-quark Polyakov line operators
\peq{L4} are taken at fixed $S^5$ angles ${\vec \theta}$ and ${\vec
  \theta}'$ correspondingly.  Using the results of
\cite{Berenstein:1998ij} we can write this contribution (somewhat
schematically) in terms of the integrals over string world-sheets
${\cal A}$ and ${\cal A}'$
\begin{align}
  \label{goal2}
  \frac{\left\langle \text{Tr} \, L^\dagger (0, {\vec \theta}') \ 
      \text{Tr} \, L ({\vec r}, {\vec \theta})
    \right\rangle}{\left\langle \text{Tr} \, L^\dagger (0, {\vec
        \theta}') \right\rangle \, \left\langle \text{Tr} \, L ({\vec
        r}, {\vec \theta}) \right\rangle} \Bigg|_{\text{adjoint}} \, =
  \, \exp \left[ \sum\limits_{k, m} Y_m^{(k)} (\theta) \, Y_m^{(k)}
    (\theta') \, \int \frac{d {\cal A}}{2 \, \pi \, \alpha'} \,
    \frac{d {\cal A}'}{2 \, \pi \, \alpha'} \, G_k (z, \tau; z',
    \tau') \right].
\end{align}
Here $G_k (z, \tau; z', \tau')$ is the sum of bulk-to-bulk propagators
(multiplied by the appropriate vertex factors different for each
field) for all the supergravity fields that couple to string
world-sheets (parametrized by $z, \tau$ and $z', \tau'$), with the
appropriate indices chosen for the graviton and 2-form contributions.
The functions $Y_m^{(k)} (\theta)$ are spherical harmonics, $Y_m^{(k)}
(\theta) \, = \, C_{I_1 \ldots I_k}^m \, \theta_{I_1} \ldots
\theta_{I_k}$ with $m$ labeling all harmonics corresponding to the
state with the total angular momentum $J^2 = k (k+4)$ and $C_{I_1
  \ldots I_k}^m$ a basis of symmetric traceless tensors such that
$C_{I_1 \ldots I_k}^{m_1} \, C^{I_1 \ldots I_k \ m_2} = \delta^{m_1 \,
  m_2}$ \cite{Lee:1998bxa,Berenstein:1998ij}. As usual the slope
parameter $\alpha' = 1/\sqrt{\lambda}$.

Expanding \eq{goal2} to the first non-trivial order, and averaging it
along with \eq{goal} over $S^5$ angles $\theta$ and $\theta'$ yields
\begin{align}
  \label{goal3}
  \left\langle V_{adj} (r) \right\rangle_{\theta} \, = \,
  \frac{1}{\beta} \, \left\langle S_{straight} \right\rangle_{\theta}
  \, = \, \frac{1}{\beta} \, \int \frac{d {\cal A}}{2 \, \pi \,
    \alpha'} \, \frac{d {\cal A}'}{2 \, \pi \, \alpha'} \, G_0 (z,
  \tau; z', \tau')
\end{align}
since only $k=0$ KK mode survives the averaging and $Y_0^0 = 1$ in the
conventions of \cite{Lee:1998bxa,Berenstein:1998ij} that we have
adopted.  Below we will calculate $k=0$ KK mode contributions to the
action $S_{straight}$ coming from the dilaton, massive scalar ($t^0$),
2-form fields, and the graviton.

%%%%%%%%%%%%%%%%%%%%%%%%%%%%%%%%%%%%%%%%%%%%%%%%%%%%%%%%%%%%%%%%%%%%%%%%%%%%

\subsection{The Dilaton}

%%%%%%%%%%%%%%%%%%%%%%%%%%%%%%%%%%%%%%%%%%%%%%%%%

\subsubsection{Dilaton potential and the EOM}

We begin with the simplest case of the dilaton exchange between the
sting world-sheets pictured in \fig{exchange}. Writing the KK
expansion for the dilaton field as $ \phi = \sum_k
\phi_{k}Y^{(k)}(\theta)$, and only taking the lowest harmonic, the
10-dimensional dilaton action will be reduced to
\begin{align}\label{dilatonaction}
S_{\phi} = %\frac{1}{2\kappa_{10}^2}\int d^{10}x\sqrt{G_{10}}~4(\nabla \phi)^2 \to
\frac{N^2_c}{16 \, \pi^2} \, \int d^5 x \, \sqrt{g}~g^{MN} \,
\partial_M\phi \, \partial_N\phi \ ,
\end{align}
where $\phi = \phi(x^{\mu},z)$ is the coefficient of the zeroth KK
harmonic and $M,N = \mu,z$ with $\mu = 0, \ldots, 3$. Notice, that the
massless dilaton in the AdS bulk is dual to the glueball operator $\Tr
F_{\mu\nu}^2$ in the boundary theory.
The coupling of the dilaton to the string world-sheet is described by
the action
\begin{align}\label{dilatonstring0}
  S_{\phi-{\rm string}} = \frac{1}{2\pi \alpha'}\int
  d^2\sigma~e^{\phi/2} \, \sqrt{\gamma} + \frac{1}{4\pi} \, \int d^2
  \sigma \, \sqrt{\gamma} \, R^{(2)} \, \phi \ ,
\end{align}
where $\gamma_{\alpha\beta} =
g_{MN}\partial_{\alpha}x^M\partial_{\beta}x^N$ is the induced metric
on the string world-sheet, and $R^{(2)}$ is the world-sheet curvature.
%% Notice, that the above action is given in the string frame, and in
%% order to compute the required amplitude, we need to go to the Einstein
%% frame, $g_{\rm str} = e^{\phi/2}g_{\rm E}$.
In the large $N_c$ limit, when the AdS radius is large, $R^{(2)}$ will
be subleading and can be ignored \cite{Berenstein:1998ij}.

Expanding $e^{\phi/2} = 1 + \phi/2 + \ldots$ and using $\alpha' =
1/\sqrt{\lambda}$ we see that the leading-order action of the dilaton
coupled to two strings is
\begin{align}\label{dilatonaction1}
  S_{{\rm dil}} \, = \, \frac{N^2_c}{16 \, \pi^2} \, \int d^5x \,
  \sqrt{g}~g^{MN} \, \partial_M\phi \, \partial_N\phi +
  \frac{\sqrt{\lambda}}{4\pi} \, \int_{(1)} \, d^2\sigma \,
  \sqrt{\gamma}~\phi + \frac{\sqrt{\lambda}}{4\pi} \, \int_{(2)} \,
  d^2\sigma \, \sqrt{\gamma}~\phi\ ,
\end{align}
where $\int_{(i)}$ has a meaning of integration over the $i^{\rm th}$
string world-sheet, $i=1,2$.  The equations of motion (EOM) for the
rescaled scalar field,
\begin{align}
  \label{resc_dil}
  \bar{\phi} \, \equiv \, \frac{N^2_c}{2 \, \pi \, \sqrt{\lambda}} \,
  \phi
\end{align}
can be written as
\begin{align}\label{dil_EOM}
  \frac{1}{\sqrt{g}} \, \partial_M \, \left[\sqrt{g} \, g^{MN} \,
    \partial_N \bar{\phi}_i \right] \, = \,
  \frac{\sqrt{\gamma}}{\sqrt{g}} \, \delta^{(3)}(\vec{x}-\vec{X}_i) \ 
  ,
\end{align}
where sub-index $i=1,2$ implies that we want to find the scalar field
created by the string localized at spatial position $\vec{X}_i$.
Choosing the quark to be located at $\vec{X}_1 = \vec r$ and the
anti-quark at $\vec{X}_2 = \vec 0$, the two strings world-sheets can
be parametrized as 
\begin{align}
  \label{strings_param}
  X_1^M = (\tau, {\vec r}, z=\sigma), \ \ \ \ \ X_2^M = (\tau,
  {\vec 0}, z = z_h - \sigma)
\end{align}
with $\tau \in [0, \beta]$ and $\sigma \in [0, z_h]$. Note that in
\eq{strings_param} the string dual to the anti-quark ($X_2$) is
oriented opposite to that of the string dual to the quark ($X_1$), as
also shown in \fig{exchange}: this will be important for the analysis
of the 2-form fields contribution.

Now, we are interested in evaluating the action \peq{dilatonaction1} on
the solution of the classical EOM \peq{dil_EOM}.  Taking into account
that $\phi = \phi_1 + \phi_2$, and dropping the terms in the action
that correspond to (anti-)quark self-energy contributions as they do
not contribute to the connected Polyakov loop correlator in \eq{goal}
that we want to find, we obtain
\begin{align}\label{dilatonaction2}
  \overline{S}_{{\rm dil}} = \frac{\sqrt{\lambda}}{8 \, \pi} \,
  \int_{(1)} d^2\sigma \sqrt{\gamma}~\phi_2 + \frac{\sqrt{\lambda}}{8
    \, \pi} \, \int_{(2)} d^2\sigma \sqrt{\gamma}~\phi_1 \ ,
\end{align}
where integration in each term is over one of the string world-sheets
of the classical field $\phi$ created by the other string. Note that
the two terms on the right-hand-side of \eq{dilatonaction2} are equal. 
The integration over string world-sheet is
\begin{align}
  \int d^2\sigma \, = \, \int\limits_0^\beta d \tau \,
  \int\limits_0^{z_h} dz.
\end{align}
The determinant of the induced metric is $\gamma = \det
\gamma_{\alpha\beta} = 1/z^4$ for the strings parametrized as in
\eq{strings_param} in the background metric \peq{AdSSBH}. Using all
this in \eq{dilatonaction2}, and noticing that in the static case
considered here the solution of classical EOM \peq{dil_EOM} is going
to be time-independent, we obtain the dilaton contribution to the
potential
\begin{align}
  \label{dil_pot}
  V_{adj}^\phi (r) \, = \, \frac{\lambda}{2 \, N_c^2} \,
  \int\limits_0^{z_h} \frac{dz}{z^2} \, \bar{\phi}_2 ({\vec r}, z).
\end{align}
As $\bar{\phi}_2$ is time-independent, we explicitly removed $\tau$
from its argument.

Since $g = \det g_{MN} = 1/z^{10}$ the EOM \peq{dil_EOM} for
$\bar{\phi}_2$ is
\begin{align}
  \label{dil_EOM2}
  z^3 \, \partial_z \left[\frac{f}{z^3} \, \partial_z \bar{\phi}_2 \right] +
  {\vec \nabla}^2 \, \bar{\phi}_2 \,  =  \,  z \, \delta^{(3)} (\vec{r}). 
\end{align}
The equation is easier to tackle in momentum space. Writing
\begin{align}
  \label{FTdil}
  \bar{\phi}_2 ({\vec r}, z) \, = \, \int \frac{d^3 q}{(2 \, \pi)^3}
  \, e^{i \, {\vec q} \cdot {\vec r}} \, \varphi ({\vec q}, z)
\end{align}
we recast \eq{dil_EOM2} as
\begin{align}
  \label{dil_EOM3}
  \left( 1 - \frac{z^4}{z^4_h} \right) \, \varphi_{z z} - \left( 3 +
    \frac{z^4}{z^4_h} \right) \, \frac{1}{z} \, \varphi_z - q^2 \,
  \varphi = z
\end{align}
with $q^2 = {\vec q}^{\ 2}$, $\varphi_{z z} = \partial_z^2 \varphi$,
$\varphi_z = \partial_z \varphi$. In solving \eq{dil_EOM3} it is
convenient to measure $z$ and $q$ in units of $z_h$, such that,
rescaling
\begin{align}
  \label{redef}
  \frac{z}{z_h} \, \rightarrow \, z, \ \ \ q \, z_h \, \rightarrow \,
  q, \ \ \ \frac{\varphi}{z_h^3} \, \rightarrow \, \varphi
\end{align}
to make them dimensionless, we get
\begin{align}
  \label{dil_EOM4}
  \left( 1 - z^4 \right) \, \varphi_{z z} - \left( 3 + z^4 \right) \,
  \frac{1}{z} \, \varphi_z - q^2 \, \varphi \, = \, z.
\end{align}
The potential in momentum space $V_{adj} (q)$ is defined by
\begin{align}
  \label{potF}
  V_{adj} (r) \, = \, \frac{1}{z_h} \, \int \frac{d^3 q}{(2 \, \pi)^3}
  \, e^{i \, {\vec q} \cdot {\vec r}} \, V_{adj} (q),
\end{align}
where we have switched to dimensionless ${\vec r}/z_h \,\rightarrow \,
{\vec r}$ as well.  The dimensionless momentum-space dilaton
contribution to the potential is
\begin{align}
  \label{pot_dil_q}
  V_{adj}^\phi (q) \, = \, \frac{\lambda}{2 \, N_c^2} \,
  \int\limits_0^{1} \frac{dz}{z^2} \, \varphi (q, z).
\end{align}
(Note that the rescaled $\varphi$ in \eq{pot_dil_q} is dimensionless,
and so are the rescaled $z$ and $q$.) Since \eq{dil_EOM4} depends on
$\vec q$ through $q^2$, its solution $\varphi$ depends on the vector
$\vec q$ through its length squared $q^2$.

%%%%%%%%%%%%%%%%%%%%%%%%%%%%%%%%%%%%%%%%%%%%%%%%%

\subsubsection{Solution of the dilaton EOM}

Solution of the dilaton EOM \peq{dil_EOM4} can be constructed as a
series in the powers of $z$
\begin{align}
  \label{dil_sol1}
  \varphi \, = \, \sum_{n=0}^\infty \, \frac{b_n \, z^{2 \, n + 3}}{2
    \, n + 3} + C^\phi (q^2) \, \sum_{n=0}^\infty \, \frac{f_n \, z^{2
      \, n + 4}}{2 \, n + 4}
\end{align}
with the recursion relations for the coefficients
\begin{subequations}
\begin{align}
  \label{b_rec}
  b_n \, = \, \frac{q^2}{(2 \, n -1) \, (2 \, n+1)} \, b_{n-1} +
  b_{n-2}, \ \ \ \ \ b_0 = -1, \ \ b_1 = - \frac{q^2}{3},
\end{align}
\begin{align}
  \label{f_rec}
  f_n \, = \, \frac{q^2}{2 \, n \, (2 \, n+2)} \, f_{n-1} +
  f_{n-2}, \ \ \ \ \ f_0 = 1, \ \ f_1 = \frac{q^2}{8}.
\end{align}
\end{subequations}
The first series in \eq{dil_sol1} is the solution of the inhomogeneous
equation, while the second series is the solution of the homogeneous
equation. Both series were constructed by requiring that $\varphi (z=0)
= 0$: this is the condition of normalizability of the dilaton field.
Clearly the potential in \peq{pot_dil_q} would be infinite if this
condition was not satisfied. The coefficient $C^\phi (q^2)$ in
\eq{dil_sol1} has to be fixed by the boundary condition at the
horizon. To obtain this condition we follow the standard procedure
and expand the AdSSBH metric \peq{AdSSBH} near horizon. Defining $\rho
= 2 \sqrt{1-z}$ we obtain a flat near-horizon metric
\begin{align}
  \label{flat}
  ds^2 \big|_{\rho \approx 0} \, \approx \, \rho^2 d \tau^2 + d \rho^2
  + d {\vec x}^{\ 2}.
\end{align}
In this new flat metric the function $\varphi (x, \rho)$ should have
zero derivative at the origin, $\varphi_\rho (\rho =0) =0$, otherwise
$\varphi$ would have a discontinuous derivative at the
origin.\footnote{We thank Samir Mathur for pointing out this argument
  to us.} This translates into
\begin{align}
  \label{zbc}
  \left\{ \sqrt{1-z} \, \varphi_z \right\} \big|_{z=1} \, = \, 0
\end{align}
in $z$-space. The condition \peq{zbc} appears to be somewhat weak, as
it only implies that $\varphi$ should be finite at the horizon. To see
that the finiteness of $\varphi$ at the horizon uniquely determines
the solution of \eq{dil_EOM4} we can expand $\varphi$ in the powers of
$1-z$. In this near-horizon expansion case the series solving the
inhomogeneous and homogeneous equations \peq{dil_EOM4} are the same,
such that the solution can be written as a single series
\begin{align}
  \label{hor_exp}
  \varphi \, = \, \sum_{n=0}^\infty \, h_n \, (1-z)^n
\end{align}
with the recursion relations
\begin{align}
  \label{h_rec}
  h_n \, = \, & \frac{2 \, (n-1) \, ( 5 \, n - 8) + q^2}{4 \, n^2} \,
  h_{n-1} - \frac{2 \, (n-2) \, (5 \, n -12) + q^2}{4 \, n^2} \,
  h_{n-2} \notag \\ & + \frac{(n-3) \, (5 \, n - 16)}{4 \, n^2} \,
  h_{n-3} - \frac{(n-4)^2}{4 \, n^2} \, h_{n-4} + \frac{1}{4} \,
  \delta_{n 1} - \frac{1}{8} \, \delta_{n 2} + \frac{1}{36} \,
  \delta_{n 3}
\end{align}
between its coefficients. One can see that the value of $\varphi$ at
the horizon $z=1$, given by the coefficient $h_0$, is assumed to be
finite here. The coefficient $h_0$ alone determines the rest of the
series \peq{hor_exp}, with the help of \eq{h_rec}. The only remaining
degree of freedom, $h_0$, is fixed by the $\varphi (z=0) = 0$
condition.  This proves that finiteness of $\varphi$ at the horizon,
along with the $\varphi (z=0) = 0$ condition, are sufficient to
uniquely define the solution of \eq{dil_EOM4}.

Analytic summation of the series \peq{dil_sol1} and/or \peq{hor_exp}
appears to be prohibitively complicated. Instead we construct the
solution for $\varphi$ by summing the series \peq{dil_sol1}
numerically. We numerically evaluate partial sums
\begin{align}
  \label{dil_sol2}
  \varphi (q, z, N) \, = \, \sum_{n=0}^N \, \frac{b_n \, z^{2 \, n + 3}}{2
    \, n + 3} + C^\phi (q^2, N) \, \sum_{n=0}^N \, \frac{f_n \, z^{2
      \, n + 4}}{2 \, n + 4}
\end{align}
constructing sequential approximations $\varphi (q, z, N)$ of the
exact solution $\varphi (q, z) = \lim\limits_{N \rightarrow \infty}
\varphi (q, z, N)$. To determine the coefficients $C^\phi (q^2, N)$ we
need to require that $\varphi (q, z, N)$ is finite at the horizon: to
insure this without specifying the (unknown) value of $\varphi (q, z,
N)$ at the horizon we impose Neumann boundary conditions on $\varphi
(q, z, N)$ at $z=1$. Indeed the exact solution of \eq{dil_EOM4} does
not have to satisfy Neumann boundary conditions at the horizon:
however, imposing this condition on partial sums $\varphi (q, z, N)$
makes them finite at the horizon, such that, since the solution which
is finite at the horizon is unique, as $N$ increases the partial sums
would converge onto this one exact solution for $\varphi (q, z)$, with
the part of the interval $z \in (0,1)$ affected by the Neumann
boundary condition rapidly shrinking. Since \eq{dil_EOM4} is a
second-order differential equation, it has two solutions: one that is
finite at the horizon, and one that is infinite at the horizon.
Therefore any boundary condition forcing $\varphi (q, z, N)$ to be
finite at $z=1$ would map these partial sums onto the solution finite
at the horizon in the $N \rightarrow \infty$ limit. For instance,
Dirichlet boundary condition at $z=1$ would also work. However,
Neumann boundary conditions appear to give the fastest numerical
convergence to the exact solution.

%%%%%%%%%%%%%%%%%%%%%%%%%%%%%%%%%%%%%%%%%%%%%%%%%%%%%%%%%%%
\FIGURE{\includegraphics[width=14cm]{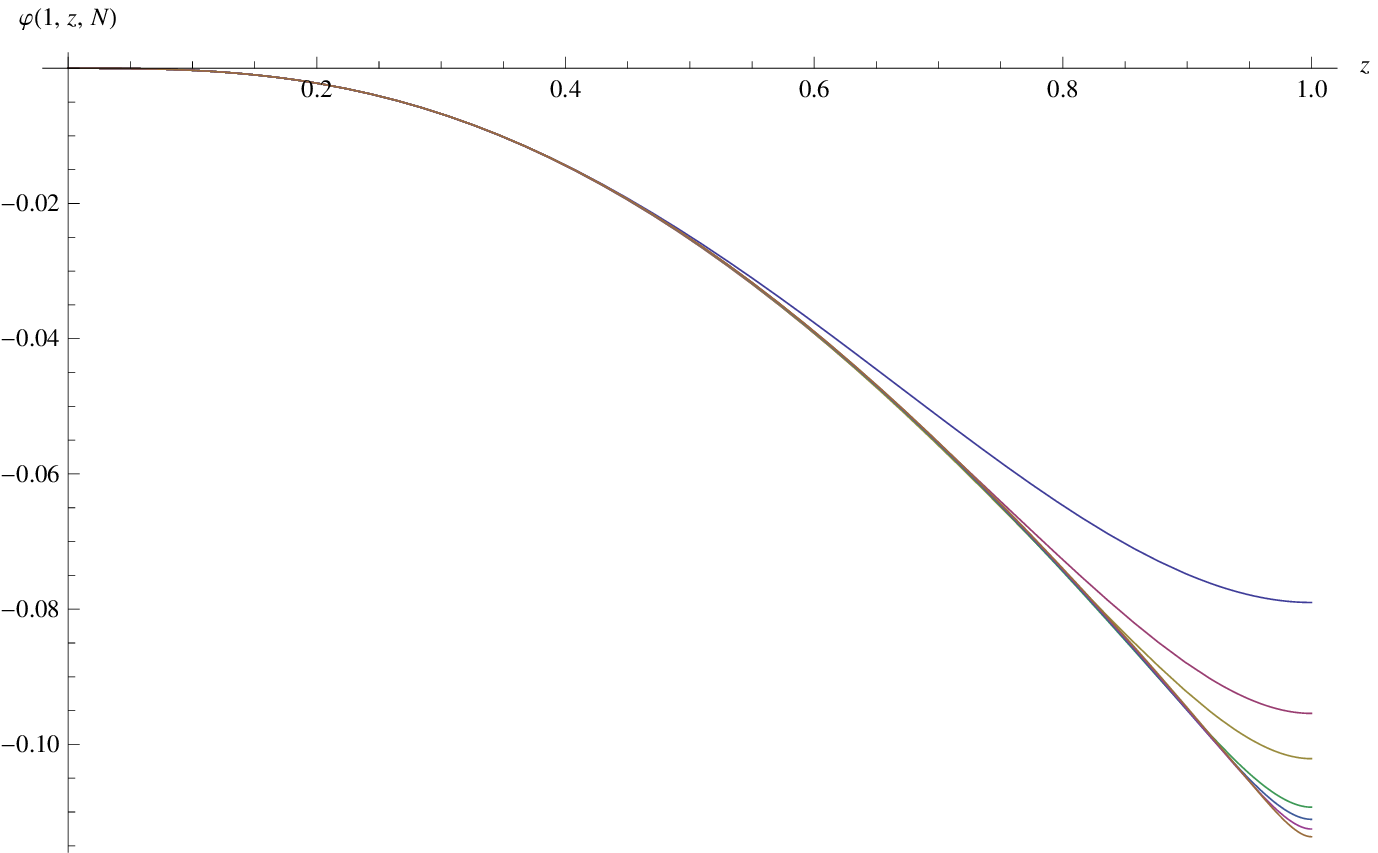}
  \caption{Numerical evaluations of $\varphi (q, z, N)$ as functions of 
    $z$ (in units of $z_h$) from \eq{dil_sol2} for $q=1$ and $N=1,3,
    5, 11, 15, 21, 31$ in the descending curve order. }
%$N=500$ (blue line), $N=1000$ (green line) and $N=5000$ (red
%    line). }
  \label{phi}
}
%%%%%%%%%%%%%%%%%%%%%%%%%%%%%%%%%%%%%%%%%%%%%%%%%%%%%%%%%%%

Demanding that $\partial_z \varphi (q, z, N) \big|_{z=1} =0$ yields
\begin{align}
  \label{CphiN}
  C^\phi (q^2, N) \, = \, - \frac{\sum_{n=0}^N \, b_n}{\sum_{n=0}^N \,
    f_n}.
\end{align}
Using \eq{CphiN} in \eq{dil_sol2} we construct a series of numerical
approximations to the solution of \eq{dil_EOM4}. Three approximate
solutions are plotted in \fig{phi} for different values of $N$. One
can see that, as the order of the partial sum $N$ increases, the
solutions tend to converge to the unified limiting curve, deviating
from it only in the rapidly shrinking region near $z=1$ to satisfy
Neumann boundary condition.

%%%%%%%%%%%%%%%%%%%%%%%%%%%%%%%%%%%%%%%%%%%%%%%%%%%%%%%%%%%
\FIGURE{\includegraphics[width=14cm]{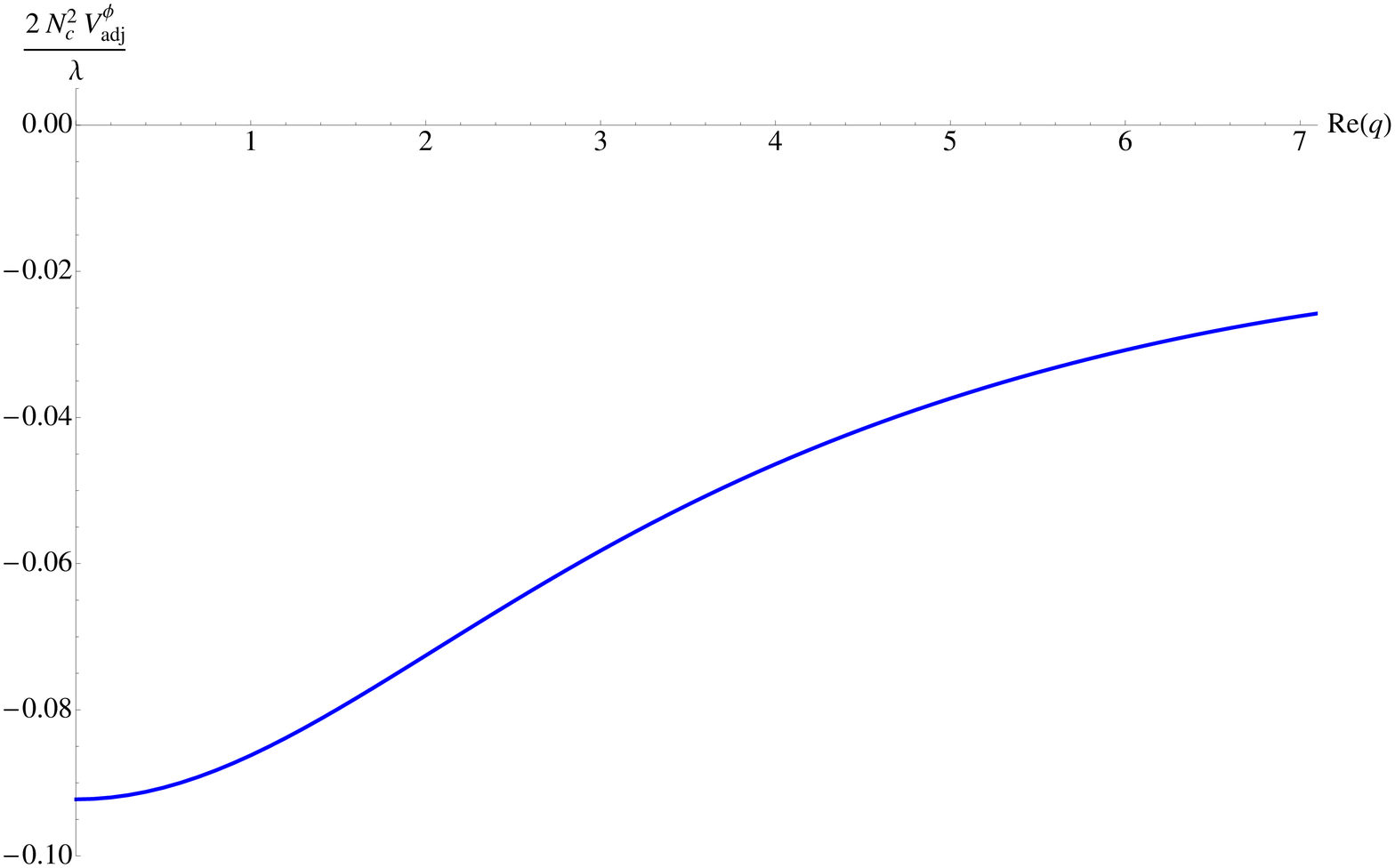}
  \caption{The contribution to the adjoint finite-$T$ $Q\bar Q$ 
    potential in momentum space due to the exchange of the dilaton
    field between the string world-sheets, plotted along the real-$q$
    axis (Im~$q=0$) in units of $\lambda/(2 \, N_c^2)$. The plot is for $N=10^4$ iterations.  (Here and in
    all subsequent plots we put $z_h =1$.)}
  \label{dilre}
}
%%%%%%%%%%%%%%%%%%%%%%%%%%%%%%%%%%%%%%%%%%%%%%%%%%%%%%%%%%%

Since we are interested in the dilaton contribution to the heavy quark
potential $V_{adj}^\phi (q)$ we 
%% define
%% \begin{align}
%%   \label{vtilde}
%%   v_{adj} (q) \, \equiv \, \frac{N_c^2}{\lambda} \, V_{adj}
%%   (q).
%% \end{align}
use \eq{dil_sol1} in \eq{pot_dil_q} to obtain
\begin{align}
  \label{dil_pot_final}
  V_{adj}^\phi (q) \, = \, \frac{\lambda}{2 \, N_c^2} \,
  \left[ \sum_{n=0}^\infty \, \frac{b_n}{(2 \, n + 2) \, (2 \, n + 3)}
    + C^\phi (q^2) \, \sum_{n=0}^\infty \, \frac{f_n}{(2 \, n + 3) \,
      (2 \, n + 4)} \right].
\end{align}
The potential is also evaluated by numerical calculation of the
partial sums. Dilaton contribution $V_{adj}^\phi (q)$ to the heavy
quark potential is plotted in momentum space in \fig{dilre}. The
potential is finite at $q=0$ which indicates screening of the
quark--anti-quark interactions by the thermal ${\cal N}=4$ SYM medium.
Since the potential is monotonically increasing and negative, one
should expect that in coordinate space the contribution of the dilaton
will be {\sl attractive}.

%%%%%%%%%%%%%%%%%%%%%%%%%%%%%%%%%%%%%%%%%%%%%%%%%

\subsubsection{Asymptotics of the dilaton contribution}

While it is hard to perform a reliable numerical Fourier transform of
the potential from \fig{dilre} into coordinate space, we can use our
numerical results to understand its asymptotics at large and small
values of the dimensionless parameter $r$, or, equivalently, of the
parameter $r \, T$ if we go back to the dimensionful $r$ not measured
in the units of $z_h$.

Let us begin with the case of large-$r \, T$, which can be achieved by
either increasing the quark--anti-quark separation $r$ or by
increasing the temperature $T$. Noting that the momentum space
potential is a function of $q^2$ we can integrate \eq{potF} over the
angles, obtaining
\begin{align}
  \label{potF1}
  V_{adj} (r) \, = \, \frac{-i}{(2 \, \pi)^2} \, \frac{1}{r \, z_h} \,
  \int\limits_{-\infty}^\infty d q \ q \, e^{i \, q \, r} \, {\tilde
    V}_{adj} (q^2).
%% \, = \, \frac{-i}{(2 \, \pi)^2} \, \frac{1}{r \,
%%     z_h} \, \frac{\lambda}{N_c^2} \, \int\limits_{-\infty}^\infty d q
%%   \ q \, e^{i \, q \, r} \, {\tilde v}_{adj} (q^2).
\end{align}
The $q$-integral can be done by closing the contour in the upper
half-plane and picking up contributions of all the singularities
there. It turns out that the singularities of the dilaton potential
$V_{adj}^\phi (q)$ (along with the similar potentials for other
supergravity fields that we will analyze below) are only limited to
poles along the imaginary-$q$ axis. As originally suggested in
\cite{Bak:2007fk}, the positions of the poles correspond to the
glueball masses for QCD$_3$ calculated in
\cite{Csaki:1998qr,Zyskin:1998tg,Brower:2000rp,Brower:1999nj,Constable:1999gb,deMelloKoch:1998qs}
and references therein. The potential $V_{adj}^\phi (q)$ is plotted in
units of $\lambda/(2 \, N_c^2)$ along the positive imaginary-$q$ axis
in \fig{dilim}, illustrating this point. Note that the potential is
real along both the real and the imaginary axes in the complex
$q$-plane.

Denoting the positions of the poles on the positive imaginary-$q$ axis
by $i \, m_n$ with $n=1, 2, \ldots$ (such that $m_1 < m_2 < \ldots$)
we write (see also \cite{Danielsson:1998wt,Zyskin:1998tg})\footnote{In
  \eq{pot_res} we assume that the poles are order one: we have
  explicitly verified this only for the first pole for each
  supergravity particle analyzed.}
\begin{align}
  \label{pot_res}
  V_{adj} (r) \, = \, \frac{1}{2 \, \pi \, r \, z_h} \,
  \sum_{n=1}^\infty i \, m_n \, e^{- m_n \, r} \, \lim_{q \,
    \rightarrow \, i \, m_n} \left[ (q - i \, m_n) \, V_{adj}
    (q^2) \right]. 
\end{align}
The large-$r$ behavior is determined by the first pole $m_1$. 

%%%%%%%%%%%%%%%%%%%%%%%%%%%%%%%%%%%%%%%%%%%%%%%%%%%%%%%%%%%
\FIGURE{\includegraphics[width=14cm]{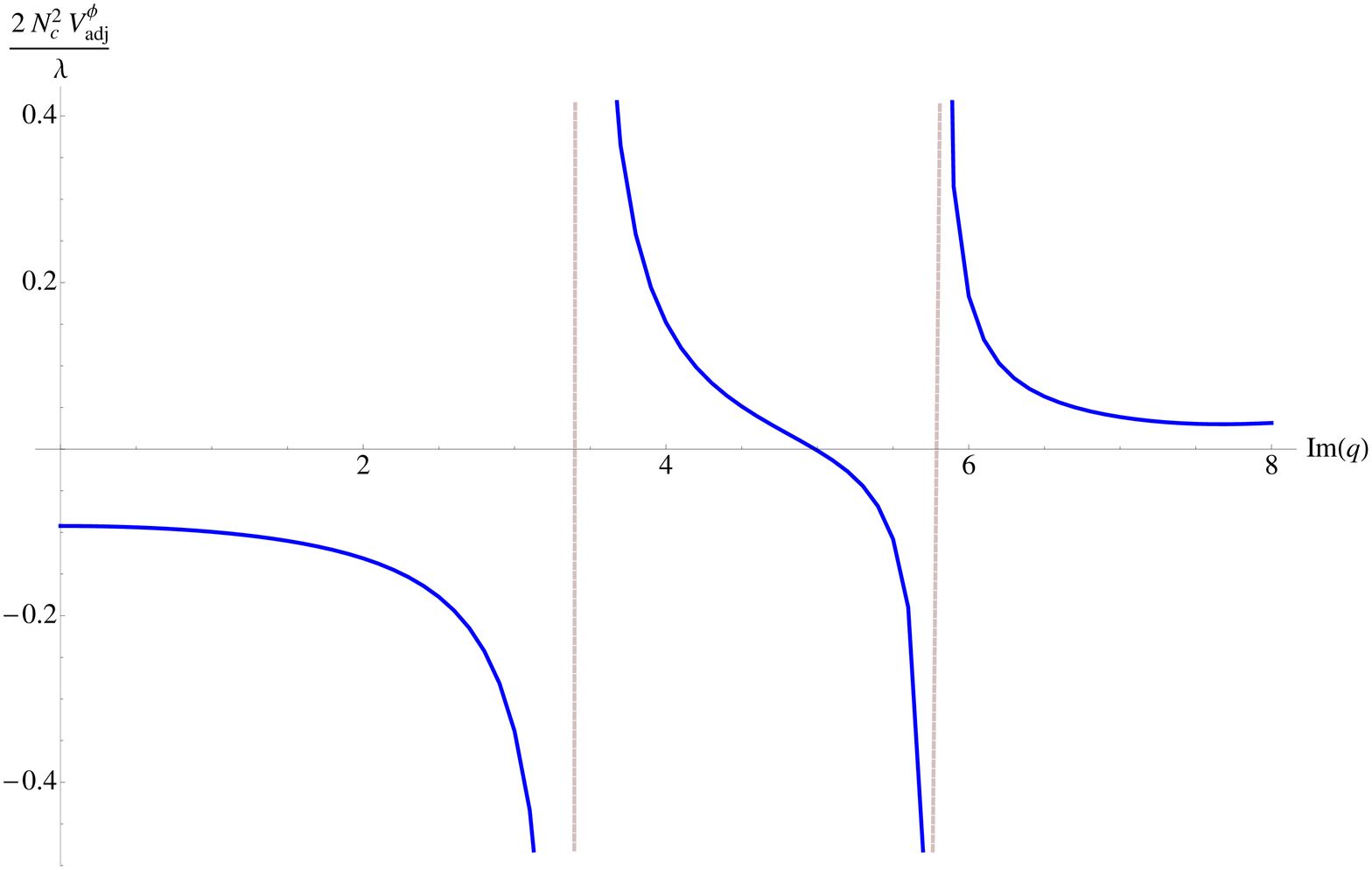}
  \caption{The contribution to the adjoint finite-$T$ $Q\bar Q$ 
    potential in momentum space due to the exchange of the dilaton
    field between the string world-sheets, plotted along the
    imaginary-$q$ axis (Re~$q=0$) in units of $\lambda/(2 \, N_c^2)$. The plot is for $N=5000$ iterations.}
  \label{dilim}
}
%%%%%%%%%%%%%%%%%%%%%%%%%%%%%%%%%%%%%%%%%%%%%%%%%%%%%%%%%%%

For the case of the dilaton, according to our numerical solution the
leading pole is at $m_1^\phi = 3.4041 \pm 0.0001$ (for $N= 3\times 10^5$ iterations), in agreement with
\cite{Csaki:1998qr,Brower:2000rp}. The residue of the pole is $i \,
(0.12 \pm 0.01)$, such that the large-$r$ asymptotics of the dilaton
contribution to the potential is
\begin{align}
  \label{large_dil}
  V_{adj}^\phi (r) \bigg|_{r \, T \gg 1} \, \approx \, -
  \frac{\lambda}{2 \, N_c^2} \, \frac{0.41 \pm 0.04}{2 \, \pi \, r} \,
  e^{- 3.4041 \, \pi \, r \, T},
\end{align}
where we have inserted $z_h = 1/(\pi \, T)$ back into the expression.
We obtain a screened Yukawa-type attractive potential falling off
exponentially with the distance $r$.

Now we consider the case of small-$r$, which, inserting $z_h$ back,
means small $r \, T$, a regime which can be interpreted as resulting
from either the short separations $r$ or low temperatures $T$. Due to
the Fourier transform \peq{potF}, low-$r$ corresponds to large-$q$. In
the large-$q$ limit, and for $z \gg 1/q$, we can neglect the terms on
the left-hand-side of \eq{dil_EOM4} containing $z$-derivatives
compared to the term with $q^2$. This yields
\begin{align}
  \label{dil_large-q}
  \varphi \big|_{q \, z \gg 1} \, \approx \, - \frac{z}{q^2}.
\end{align}
Since, as follows from the series solution \peq{dil_sol1}, the
function $\varphi$ goes to zero as $z^3$ when $z \rightarrow 0$, we
can argue that most of the support of the integrand $\varphi/z^2$ in
\eq{pot_dil_q} comes from larger $z$, and hence, for large-$q$, the
condition $q \, z \gg 1$ is satisfied over most of the $z$-range
contributing to the potential. Using \eq{dil_large-q} in
\eq{pot_dil_q} and inserting a UV cutoff of $1/q$ in it, we obtain
(with the leading logarithmic accuracy)
\begin{align}
  \label{Vdil-large-q}
  V_{adj}^\phi (q) \bigg|_{q \gg 1} \, \approx \, -
  \frac{\lambda}{2 \, N_c^2} \, \frac{1}{q^2} \, \ln q.
\end{align}
Fourier-transforming this back into coordinate space with the help of
\eq{potF}, and inserting $z_h$ back in to make $r$ dimensionful yields
(again with the leading logarithmic accuracy)
\begin{align}
  \label{Vdil-small-r}
  V_{adj}^\phi (r) \bigg|_{r \, T \ll 1} \, \approx \, -
  \frac{\lambda}{2 \, N_c^2} \, \frac{1}{4 \, \pi \, r} \, \ln
  \frac{1}{r \, T}.
\end{align}
The potential in \peq{Vdil-small-r} is attractive, but is not quite
Coulomb-like, due to the extra logarithmic factor. This factor is
somewhat worrisome, as it makes the potential infinite in the $T
\rightarrow 0$ limit. This, however, might not be viewed as a problem
on the gauge theory side, since it is not clear how to define the
adjoint potential at $T=0$ in a gauge-invariant way: therefore, the
divergence takes place in the limit when the quantity we are
calculating is not defined.  On the string theory side it appears
that, in absence of the black hole, the interaction between two string
world-sheets would be infinite: the situation is analogous to the
instability of the parallel D-brane--anti-D-brane configuration. One
may also argue that at small separations $r$ higher-order string
corrections may come in and modify the result, possibly regulating the
divergence: we will discuss this possibility after calculating the
contributions of all other supergravity fields.

%%%%%%%%%%%%%%%%%%%%%%%%%%%%%%%%%%%%%%%%%%%%%%%%%%%%%%%%%%%%%%%%%%%%%%%%%%%%%%

\subsection{Massive Scalars}

%%%%%%%%%%%%%%%%%%%%%%%%%%%%%%%%%%%%%%%%%%%

\subsubsection{The potential and EOM}

In addition to the dilaton, compactification on $S^5$ leads to
additional scalar modes that come from the KK modes of the trace of
the metric and from the 4-form field perturbations over $S^5$, that
is, from $h^{\alpha}_{\alpha}$ and $a_{\alpha\beta\gamma\delta}$
correspondingly, where indices take values on $S^5$
\cite{Kim:1985ez,Lee:1998bxa,Berenstein:1998ij,Arutyunov:1999en,Arutyunov:1999fb}.
The KK modes of $h^{\alpha}_{\alpha}$ and
$a_{\alpha\beta\gamma\delta}$ mix in the EOM, but at the quadratic
order the action can be diagonalized by the scalar modes $s^k$ (with
$k \ge 2$) and $t^k$ (with $k \ge 0$)
\cite{Kim:1985ez,Lee:1998bxa,Berenstein:1998ij,Arutyunov:1999en,Arutyunov:1999fb}.
The modes $s^k$ for $k=2,3$ are tachyonic, while all other $s^k$ and
all $t^k$ modes are massive: we will refer to all modes $t^k$ and
$s^k$ as massive scalars.
%: here we will follow the
%notation of \cite{Berenstein:1998ij} and call all modes $t^k$ and
%$s^k$ tachyonic. 
Since we are interested in the modes which survive averaging over
$S^5$, i.e., in $k=0$ KK modes, we only need to study the contribution
of $t^0$ to the string interaction. The action for the $t^k$ modes can
be found in \cite{Arutyunov:1999en,Arutyunov:1999fb}.  The coupling of
$t^k$ to the string world-sheet follows from the coupling of the
graviton to the string. Writing the 5d metric fluctuations for $k=0$
KK mode in terms of fields diagonalizing EOM as
\cite{Arutyunov:1999fb}
\begin{align}
  \label{Weyl_shift}
  \delta g_{MN}^0 \, = \, h_{MN}^0 - \frac{40}{3} \, g_{MN} \, t^0 +
  \frac{4}{3} \, \nabla_M \, \nabla_N \, t^0
\end{align}
with $\nabla_M$ the covariant derivative we can determine the coupling
of $t^0$ to the string world-sheet from the coupling of $\delta
g_{MN}^0$ to the string. One can readily show that the coupling of the
$\nabla_M \, \nabla_N \, t^0$ term to the world-sheet of our straight
strings \peq{strings_param} is zero if $\partial_z t^0 (z=0) =0$, with
the latter condition satisfied by the solution of the classical EOM
for $t^0$. Dropping the derivative term in \eq{Weyl_shift} we get
(using the same notation as in \eq{dilatonaction1})
\begin{align}
 \label{t0action1}
 S_{t^0} \, = \, \frac{80}{3} \, \frac{N^2_c}{\pi^2} \, \int d^5x \,
 \sqrt{g} \, \left[ g^{MN} \, \partial_M t^0 \, \partial_N t^0 + 32
   (t^0)^2 \right] + \frac{20}{3} \, \frac{\sqrt{\lambda}}{\pi} \,
 \int\limits_{(1)} \, \frac{d \tau \, dz}{z^2} \ t^0 + \frac{20}{3} \,
 \frac{\sqrt{\lambda}}{\pi} \, \int\limits_{(2)} \, \frac{d \tau \,
   dz}{z^2} \, t^0 .
\end{align}
Rescaling the $t^0$ field
\begin{align}
  \label{resc_t0}
  \bar{t}^{\ 0} \, \equiv \, \frac{8 \, N^2_c}{\pi \, \sqrt{\lambda}}
  \, t^0
\end{align}
we write the equation of motion for the field produced by the string
at ${\vec X}_2 = {\vec 0}$:
\begin{align}\label{t0_EOM}
  \frac{1}{\sqrt{g}} \, \partial_M \, \left[\sqrt{g} \, g^{MN} \,
    \partial_N \bar{t}^{\ 0} \right] - 32 \, \bar{t}^{\ 0} \, = \, z^3
  \, \delta^{(3)}(\vec{x}) \ .
\end{align}
The action evaluated at the classical solution is
\begin{align}
 \label{t0action2}
 {\overline S}_{t^0} \, = \, \frac{5}{6} \, \frac{\lambda}{N_c^2} \,
 \int\limits_0^\beta d \tau \, \int\limits_0^{z_h} \frac{dz}{z^2} \ 
 \bar{t}^{\ 0} ({\vec r}, z)
\end{align}
giving the contribution to the heavy quark potential
\begin{align}
 \label{t0_pot}
 V^{t^0}_{adj} (r) \, = \, \frac{5}{6} \, \frac{\lambda}{N_c^2} \,
 \int\limits_0^{z_h} \frac{dz}{z^2} \, \bar{t}^{\ 0} ({\vec r}, z).
\end{align}

Just like with the dilaton, in order to solve \eq{t0_EOM} we go to
momentum space
\begin{align}
  \label{FTt0}
  \bar{t}^0 ({\vec r}, z) \, = \, \int \frac{d^3 q}{(2 \, \pi)^3} \,
  e^{i \, {\vec q} \cdot {\vec r}} \, \tgt^0 ({\vec q}, z),
\end{align}
where, in units of $z_h$ \peq{redef}, the EOM becomes
\begin{align}
  \label{t0_EOM1}
  \left( 1 - z^4 \right) \, \tgt^0_{z z} - \left( 3 + z^4
  \right) \, \frac{1}{z} \ \tgt^0_z - q^2 \, \tgt^0 \,
  - \frac{32}{z^2} \, \tgt^0 \, = \, z.
\end{align}
Just like with the dilaton we require that $\tgt^0 (z=0) =0$
and $\tgt^0 (z=1)$ is finite. The contribution of $t^0$ to the
momentum-space potential is then (in the same $z_h=1$ units)
\begin{align}
  \label{pot_t0_q}
  V_{adj}^{t^0} (q) \, = \, \frac{5}{6} \,
  \frac{\lambda}{N_c^2} \, \int\limits_0^{1} \frac{dz}{z^2} \ 
  \tgt^0 (q, z).
\end{align}
The coordinate-space potential can be obtained from \eq{potF}.

%%%%%%%%%%%%%%%%%%%%%%%%%%%%%%%%%%%%%%%%%%%

\subsubsection{Solution of the EOM for $t^0$}

The series solution for \eq{t0_EOM1} is
\begin{align}
  \label{t0_sol1}
  \tgt^0 \, = \, \sum_{n=0}^\infty \, c_n \, z^{2 \, n + 3} + C^{t^0}
  (q^2) \, \sum_{n=0}^\infty \, g_n \, z^{2 \, n + 8}
\end{align}
with the recursion relations for the coefficients
\begin{subequations}
\begin{align}
  \label{c_rec}
  c_n \, = \, \frac{q^2}{(2 \, n -5) \, (2 \, n+7)} \, c_{n-1} +
  \frac{(2 \, n -1)^2}{(2 \, n -5) \, (2 \, n+7)} \, c_{n-2}, \ \ \ \ 
  \ c_0 = - \frac{1}{35}, \ \ c_1 = \frac{q^2}{945},
\end{align}
\begin{align}
  \label{g_rec}
  g_n \, = \, \frac{q^2}{4 \, n \, (n+6)} \, g_{n-1} +
  \frac{(n+2)^2}{n \, (n+6)} \, g_{n-2}, \ \ \ \ \ g_0 = 1, \ \ g_1 =
  \frac{q^2}{28}.
\end{align}
\end{subequations}
Again we sum the series \peq{t0_sol1} numerically, imposing the
Neumann boundary conditions at the horizon. These yield
\begin{align}
  \label{Ct0N}
  C^{t^0} (q^2, N) \, = \, - \frac{\sum_{n=0}^N \, c_n \, (2 \, n
  +3)}{\sum_{n=0}^N \, g_n \, 2 \, (n+4)}.
\end{align}

%%%%%%%%%%%%%%%%%%%%%%%%%%%%%%%%%%%%%%%%%%%%%%%%%%%%%%%%%%%
\FIGURE{\includegraphics[width=12cm]{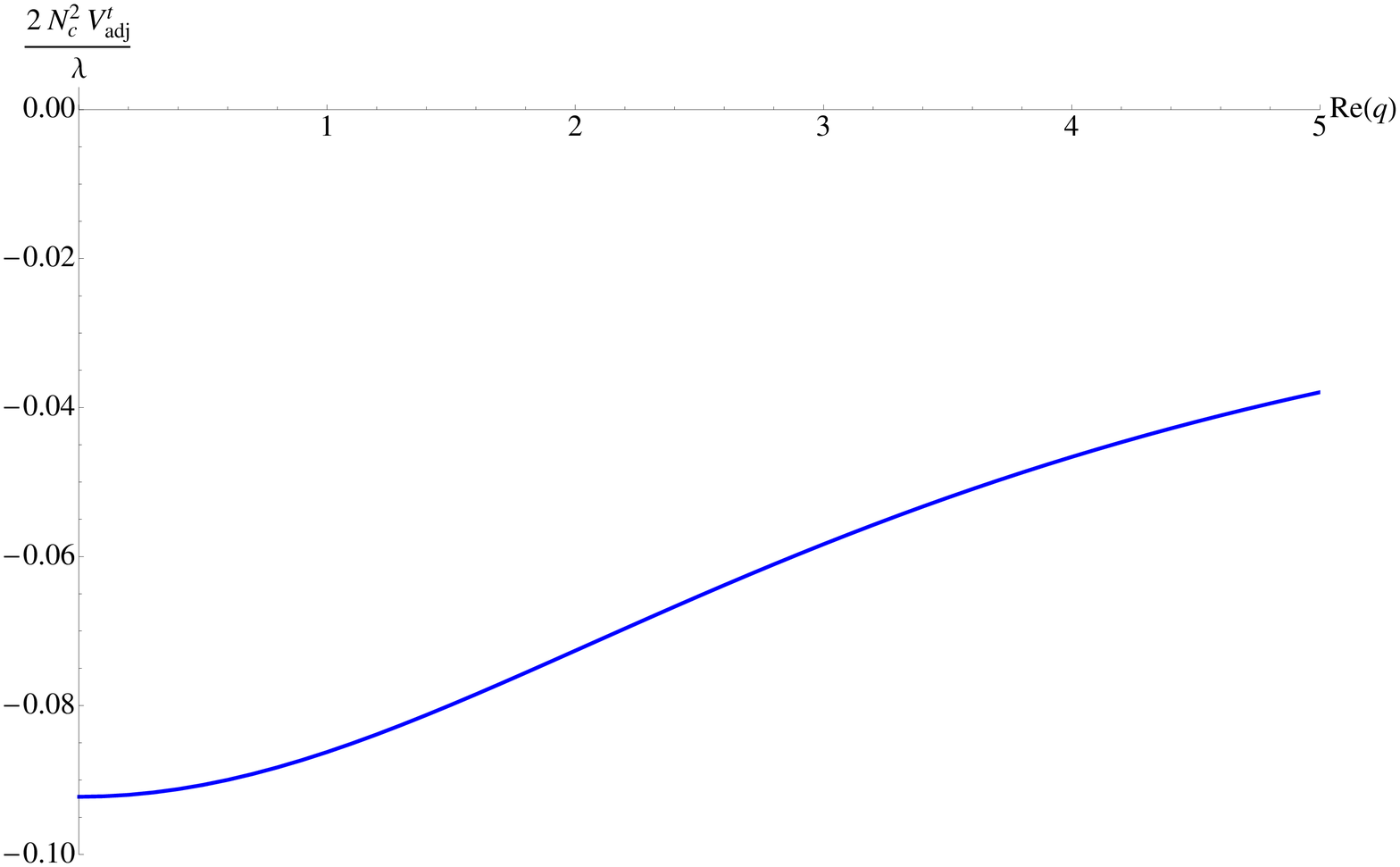}
  \caption{The contribution to the adjoint finite-$T$ $Q\bar Q$ 
    potential in momentum space due to the exchange of the $t^0$ field
    between the string world-sheets, plotted along the real-$q$ axis
    (Im~$q=0$) in units of $\lambda/(2 \, N_c^2)$. The plot is for $N=10^5$ iterations.}
  \label{t0re}
}
%%%%%%%%%%%%%%%%%%%%%%%%%%%%%%%%%%%%%%%%%%%%%%%%%%%%%%%%%%%

Substituting \eq{t0_sol1} into \eq{pot_t0_q} we obtain the
contribution of the scalar $t^0$ to the momentum-space heavy quark
potential
\begin{align}
  \label{pot_t0_q2}
  V_{adj}^{t^0} (q) \, = \, \frac{5}{6} \,
  \frac{\lambda}{N_c^2} \, \left[ \sum_{n=0}^\infty \, \frac{c_n}{2 \,
      (n+1)} + C^{t^0} (q^2) \, \sum_{n=0}^\infty \, \frac{g_n}{2 \, n
      +7} \right]. 
\end{align}
The potential $V_{adj}^{t^0} (q)$ is plotted in \fig{t0re} in
units of $\lambda/(2 N_c^2)$. It looks similar to the dilaton
contribution to the potential plotted in \fig{dilre} and also appears
to be {\sl attractive}.

%%%%%%%%%%%%%%%%%%%%%%%%%%%%%%%%%%%%%%%%%%%%%%%%%

\subsubsection{Asymptotics of the massive scalar contribution}

Just like for the dilaton, the singularities of the potential ${\tilde
  V}_{adj}^{t^0} (q)$ are poles along the imaginary-$q$ axis. The
potential $V_{adj}^{t^0} (q)$ along the positive
imaginary-$q$ axis is plotted in \fig{t0im}.

%%%%%%%%%%%%%%%%%%%%%%%%%%%%%%%%%%%%%%%%%%%%%%%%%%%%%%%%%%%
\FIGURE{\includegraphics[width=12cm]{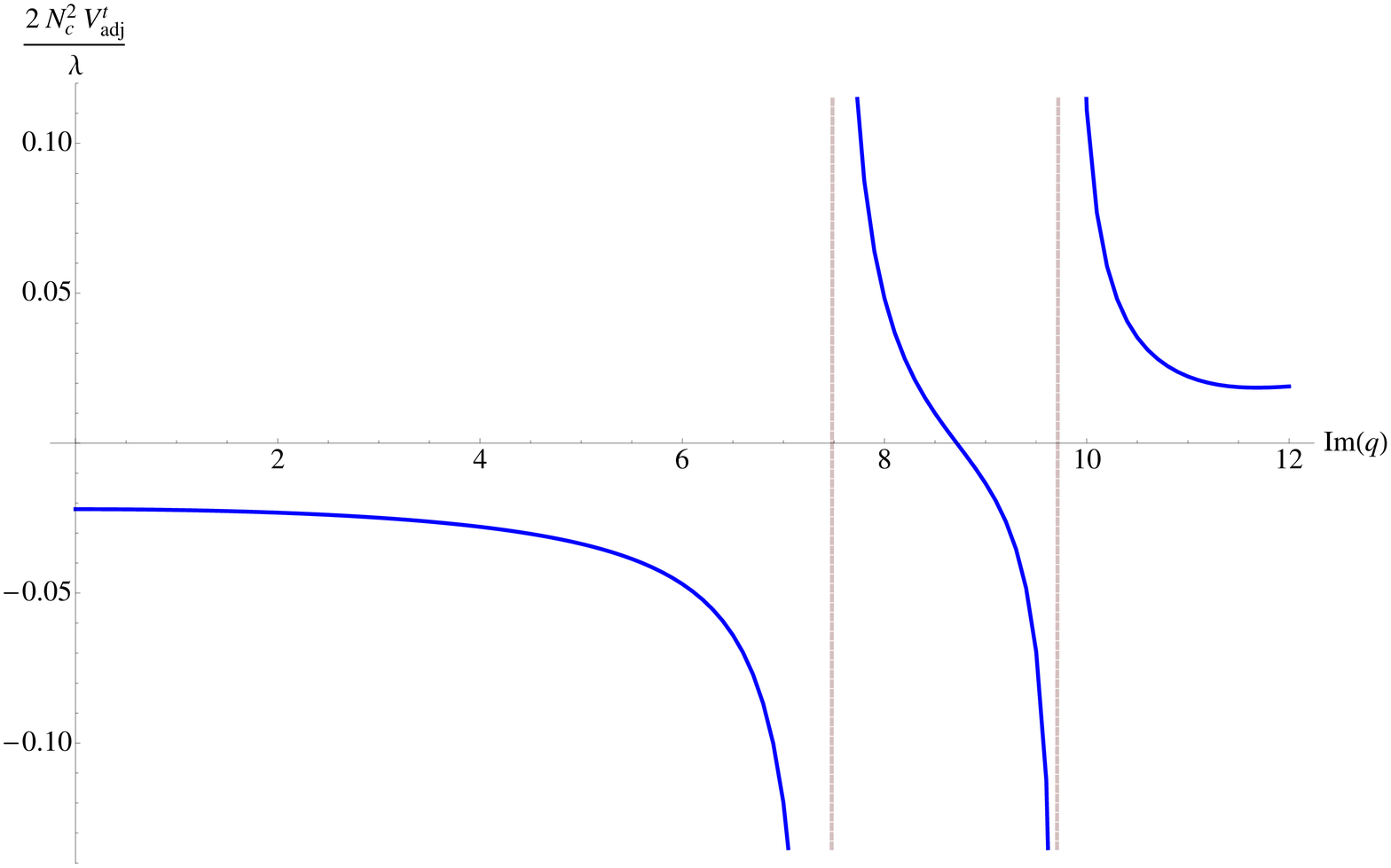}
  \caption{The contribution to the adjoint finite-$T$ $Q\bar Q$ 
    potential in momentum space due to the exchange of the $t^0$ field
    between the string world-sheets, plotted along the imaginary-$q$
    axis (Re~$q=0$) in units of $\lambda/(2 \, N_c^2)$. The plot is for $N=10^4$ iterations.}
  \label{t0im}
}
%%%%%%%%%%%%%%%%%%%%%%%%%%%%%%%%%%%%%%%%%%%%%%%%%%%%%%%%%%%

The large-$r$ asymptotics is given by the leading pole, which was
found to be at $m_1^{t^0} = 7.410 \pm 0.001$ (for $N= 10^6$ iterations), in minor disagreement
with \cite{Brower:2000rp}, possibly due to differences in
implementation of the numerical simulations. The residue of the pole
is $i \, (0.04 \pm 0.01)$, giving the large-$r$ asymptotics of the
$t^0$ contribution to the heavy quark potential
\begin{align}
  \label{pot_t0_rlarge}
  V_{adj}^{t^0} (r) \bigg|_{r \, T \gg 1} \, \approx \, - \frac{5}{6}
  \, \frac{\lambda}{N_c^2} \, \frac{0.30 \pm 0.08}{2 \, \pi \, r} \,
  \, e^{- 7.410 \, \pi \, r \, T}
\end{align}
with $z_h = 1/(\pi \, T)$ inserted back such that $r$ has units of
distance. Again, this is an attractive screened Yukawa-type potential,
with the screening mass larger than that for the dilaton. Hence the
dilaton contribution dominates over that of $t^0$ at large-$r \, T$.

We have also investigated the contribution of the tachyonic scalar
$s^2 \equiv s^{k=2}$ to the interaction between the string
world-sheets. Since this is a $k=2$ KK mode, it does not contribute to
the potential \peq{goal3} averaged over $S^5$ positions of both
strings that we want to calculate. In the effort to translate the
physical meaning of the $s^2$ contribution into the QCD language, and
identifying the $SU (4)$ $R$-symmetry group of ${\cal N} =4$ SYM with
the flavor group on QCD, one may argue that $s^2$-exchange would
correspond to some flavor-changing interaction between the quarks.
Our motivation for studying $s^2$ lies in the fact it is a tachyon,
and it is precisely the $s^2$ contribution which dominates the
large-$r$ asymptotics of the interactions of two strings in the empty
AdS$_5$ space (zero-temperature case in the gauge theory), as was
shown in \cite{Berenstein:1998ij}, giving a potential between two
rectangular Wilson loops (two ``mesons'') that falls of as $1/r^4$ at
large-$r$. The $s^2$-exchange in the AdSSBH background leads to an
attractive potential, but with a small screening mass $m_1^{s^2}
\approx 1.373 \pm 0.001$ (for $N=10^5$ iterations), such that
\begin{align}
  \label{pot_s2_rlarge}
  V_{adj}^{s^2} (r) \bigg|_{r \, T \gg 1} \, \propto \, -
  \frac{\lambda}{N_c^2} \, \frac{1}{r} \, \, e^{- 1.373 \, \pi \, r \,
    T}.
\end{align}
Clearly, if we allow for exchanges of higher KK modes, the $s^2$
contribution would dominate over that of the dilaton and $t^0$ at
large-$r \, T$. We would like to point out in advance that the
screening masses for the 2-form field and the graviton ($k=0$ term
only) would also be larger than $m_1^{s^2} \approx 1.373$. As the
screening mass appears to increase with $k$, one may conjecture that
the $s^2$ contribution \peq{pot_s2_rlarge} would dominate over all
other supergravity modes exchanges at large-$r \, T$.  Since our main
analysis here is limited to $k=0$ case when quarks do not carry
$R$-charge, we will not have the $s^2$ contribution in the net
potential.

At small-$r$/large-$q$ the approximate solution of \eq{t0_EOM1} is
\begin{align}
  \label{t0_large-q}
  \tgt^0 \big|_{q \, z \gg 1} \, \approx \, - \frac{z}{q^2},
\end{align}
giving the coordinate-space potential
\begin{align}
  \label{Vt0-small-r}
  V_{adj}^{t^0} (r) \bigg|_{r \, T \ll 1} \, \approx \, - \frac{5}{6}
  \, \frac{\lambda}{N_c^2} \, \frac{1}{4 \, \pi \, r} \, \ln
  \frac{1}{r \, T},
\end{align}
where again we have reinstated the temperature $T$ explicitly.

%%%%%%%%%%%%%%%%%%%%%%%%%%%%%%%%%%%%%%%%%%%%%%%%%%%%%%%%%%%%%%%%%%%%%%%%%%%

\subsection{Two-Form Fields}

%%%%%%%%%%%%%%%%%%%%%%%%%%%%%%%%%%%%%%%%%%%

\subsubsection{The potential and EOM}

The effective action for the $k=0$ KK mode of the NS 2-form along with
its coupling to the string world-sheet can be written as (after
integrating out the contribution of the RR 2-form)
\begin{align}\label{KRaction}
  S_{B} = & \frac{1}{24 \, \kappa^2_5} \, \int d^5x\sqrt{g}~H_{MNR} \,
  H^{MNR} + \frac{2}{\kappa^2_5} \, \int d^5x \sqrt{g}~B_{MN}B^{MN}
  \notag \\ & + \frac{\xi}{4\pi\alpha'} \, \epsilon^{ab} \,
  \int\limits_{(1)} d^2\sigma~B_{MN} \, \partial_aX^M \, \partial_bX^N
  + \frac{\xi}{4\pi\alpha'} \, \epsilon^{ab} \, \int\limits_{(2)}
  d^2\sigma~B_{MN} \, \partial_aX^M \, \partial_b X^N \ ,
\end{align}
where indices $a,b = 0,1$ run over string world-sheets, $\epsilon^{ab}
= -\epsilon^{ba}$ with $\epsilon^{01} = 1$, and $\kappa_5^2 =
4\pi^2/N^2_c$. As usual
\begin{align}
  \label{Hmnr}
  H_{MNR} \, = \, \partial_M \, B_{NR} + \partial_N \, B_{RM} +
  \partial_R \, B_{MN}.
\end{align}
The coefficient $\xi$ is equal to $1$ when both indices $M, N$ of
$B_{MN}$ in the same term are spatial, i.e., when $M, N = \mu, \nu =
1, \ldots , 4$, and $\xi = i$ when either $M$ or $N$ is $0$. This
extra factor of $i$ arises due to the transition to
Euclidean-signature space with $B_{0 \mu}^{\text{Minkowskian}} = i \,
B_{0 \mu}^{\text{Euclidean}}$, where $\mu = 1, \ldots , 4$ now.

The second term on the right of \eq{KRaction} describes the effective
mass of the 2-form that is generated via the Chern--Simons type of
mixing with the RR 2-form $C_{MN}$ \cite{Kim:1985ez,Brower:2000rp}.
To understand why there should be a mass term, notice that, in the
absence of the string sources, the 2-forms $B_{MN}$ and $C_{MN}$ can
be combined into one complex 2-form field $A_{MN} = B_{MN} + i \,
C_{MN}$ \cite{Kim:1985ez,Brower:2000rp}.  The field equation for
$A_{MN}$ can be factorized into two first order differential
equations, and each can be iterated leading to a second order equation
of the form $Max \, A_{MN} - m^2 \, A_{MN} = 0$, where $Max$ is a
second order Maxwell differential operator for 2-forms defined by $Max
\, B_{MN} = \nabla^R \, H_{MNR}$ (for more details see
\cite{Brower:2000rp}). For $k=0$ KK modes that have no dependence on
$S^5$ coordinates, one equation has $m^2 = 16$, and the other one has
$m=0$.  It can be shown that the massless equation has only pure gauge
solutions, which can be ignored (see \cite{Kim:1985ez}), leaving us
with the massive equation for the combined field $A_{MN}$. Taking real
and imaginary parts of this equation (working for the moment in the
Lorentzian-signature metric where both $B_{MN}$ and $C_{MN}$ are real)
we would get separate differential equation for $B_{MN}$ and $C_{MN}$,
namely $Max \, B_{MN} - 16 \, B_{MN} = 0$ and $Max \, C_{MN} - 16 \ 
C_{MN} = 0$. Since only the NS 2-form field $B_{MN}$ couples to string
world-sheets, we are interested in the resulting action for it that
gives this massive EOM, which is given by the first two terms on the
right of \peq{KRaction}.

Using the string parametrization \peq{strings_param} in the action
\peq{KRaction} simplifies it to
\begin{align}\label{KRaction1}
  S_{B} = & \frac{N_c^2}{96 \, \pi^2} \, \int d^5x\sqrt{g}~H_{MNR} \,
  H^{MNR} + \frac{N_c^2}{2 \, \pi^2} \, \int d^5x
  \sqrt{g}~B_{MN}B^{MN} \notag \\ & + \frac{i \, \sqrt{\lambda}}{2 \,
    \pi} \, \int\limits_{(1)} d \tau \, dz \, B_{0z} (\tau, {\vec r},
  z) - \frac{i \, \sqrt{\lambda}}{2 \, \pi} \, \int\limits_{(2)} d
  \tau \, dz \, B_{0z} (\tau, {\vec 0}, z),
\end{align}
with the action along the classical solution
\begin{align}
  \label{KRaction_cl}
  {\overline S}_{B} = \frac{i \, \sqrt{\lambda}}{2 \, \pi} \,
  \int\limits_0^\beta d \tau \, \int\limits_0^{z_h} dz \, B_{0z}
  (\tau, {\vec r}, z).
\end{align}
To find the classical field $B_{MN} (\tau, {\vec r}, z)$ due to the
string $X_2$ we define the rescaled 2-form field
\begin{align}
  \label{B_resc}
  \bar{B}_{MN} \, \equiv \, i \, \frac{N_c^2}{4 \, \pi \,
    \sqrt{\lambda}} \, B_{MN}. 
\end{align}
We need to find the classical field $\bar{B}_{MN}$ satisfying the EOM
which follow from \eq{KRaction}
\begin{align}\label{KR_EOM1}
  \frac{1}{\sqrt{g}} \, g_{M M'} \, g_{N N'} \, \partial_{R} \,
  \left[\sqrt{g} \, g^{M' P} \, g^{N' Q} \, g^{R S} \, \bar{H}_{PQS}
  \right] - 16 \, \bar{B}_{M N} = \frac{1}{\sqrt{g}} \,
  \delta^{(3)}(\vec{x}) \, \left(g_{M 0} \, g_{N z} - g_{N 0} \, g_{M
      z}\right).
\end{align}
The solution of this equation, along with \eq{KRaction_cl}, would give
us the contribution of the 2-form field to the adjoint $Q\bar Q$
potential
\begin{align}
  \label{KR_pot_resc}
  V_{adj}^B (r) \, = \, \frac{2 \, \lambda}{N_c^2} \,
  \int\limits_0^{z_h} dz \, \bar{B}_{0z} (r,z).
\end{align}

To find $\bar{B}_{0z}$ we need to solve the EOM \peq{KR_EOM1} to find
the field of the string $X_2$. In the static case we consider, and due
to rotational $O(3)$ symmetry around the string in the $R^3$-space
spanned by $x^1, x^2, x^3$, the solution should depend on $r$ and $z$
only, $\bar{B}_{MN} = \bar{B}_{MN} (r,z)$. Working in spherical
coordinates $r, \theta, \phi$ in $R^3$ we notice that EOM for
$\bar{B}_{MN}$ components with either (or both) $M$ or $N$ equal
$\theta$ or $\phi$ decouple, and since there is no source for those
components we can put them all to zero. This leaves us with
$\bar{B}_{0z}$, $\bar{B}_{0r}$, and $\bar{B}_{rz}$. The $r \, z$
component of \eq{KR_EOM1} requires that $\bar{B}_{rz} =0$, leaving us
with $\bar{B}_{0z}$ and $\bar{B}_{0r}$ only.

The EOM for components $\bar{B}_{0z}$ and $\bar{B}_{0r}$ are mixed
with each other. The $0 \, z$ and $0 \, r$ components of \eq{KR_EOM1}
yield (keeping in mind that the fields depend only on $(r,z)$)
\begin{subequations}\label{KR_EOM2}
\begin{align}
  (0 \, z) \hspace*{1cm} &\frac{z^2}{r^2} \, \partial_{r} \left[r^2
    \left( \partial_{r} \, \bar{B}_{0z} - \partial_z \, \bar{B}_{0r}
    \right) \right] - 16 \, \bar{B}_{0z}
  = z \, \delta^3 ({\vec r}) \ , \\
  (0 \, r) \hspace*{1cm} & z \, f (z) \, \partial_z \left[ z \left(
      \partial_r \, \bar{B}_{0z} - \partial_z \, \bar{B}_{0r} \right)
  \right] + 16 \, \bar{B}_{0r} = 0 \ .
\end{align}
\end{subequations}
Defining $\Lambda (r,z)$ by $\partial_r \, \Lambda = \bar{B}_{0r}$ we
can shift the 2-form field
\begin{align}
  \label{Bshift}
  \tilde{B}_{0z} \, = \, \bar{B}_{0z} - \partial_z \, \Lambda
\end{align}
and use the new field $\tilde{B}_{0z}$ to simplify Eqs.~\peq{KR_EOM2}
to
\begin{subequations}\label{KR_EOM3}
\begin{align}
  &\frac{z^2}{r^2} \, \partial_{r} \left[r^2 \partial_{r} \,
    \tilde{B}_{0z} \right] - 16 \, \left( \tilde{B}_{0z} + \partial_z
    \, \Lambda \right)  = z \, \delta^3 ({\vec r}) \ , \\
  &z \, f (z) \, \partial_z \left[ z \, \tilde{B}_{0z} \right] + 16 \,
  \Lambda = 0 \ .\label{KR_EOM3b}
\end{align}
\end{subequations}
Fourier-transforming the 2-form field
\begin{align}
  \label{FTB}
  \tilde{B}_{0z} ({\vec r}, z) \, = \, \int \frac{d^3 q}{(2 \, \pi)^3}
  \, e^{i \, {\vec q} \cdot {\vec r}} \ {\cal B}_{0z} ({\vec q}, z)
\end{align}
and eliminating $\Lambda$ from Eqs.~\peq{KR_EOM3} yields
\begin{align}\label{KR_EOM4}
  z^2 \, (1 - z^4) \, {\cal B}''_{0z} + z \, (3 - 7 \, z^4) \, {\cal
    B}'_{0z} - (15 + 5 \, z^4 + q^2 \, z^2) \, {\cal B}_{0z} = z \ .
\end{align}
Here the prime denotes the partial derivative with respect to $z$. We
have also switched to the units of $z_h$ (see \eq{redef}) in
\eq{KR_EOM4} with ${\cal B}_{0z}/z_h \rightarrow {\cal B}_{0z}$. The
boundary conditions for ${\cal B}_{0z}$ are the same as before: ${\cal
  B}_{0z} (z=0) =0$ and ${\cal B}_{0z} (z=1)$ should be finite.

Note that \eq{KR_EOM3b} insures that $\Lambda (z=0) = \Lambda (z=z_h)
=0$. Together with \eq{Bshift} these conditions allow one to replace
$\bar{B}_{0z}$ in \eq{KR_pot_resc} with $\tilde{B}_{0z}$ without
changing the value of the integral. In momentum space the contribution
of the 2-form field to the adjoint $Q\bar Q$ potential (with the
Fourier transform defined by \eq{potF}) is
\begin{align}
  \label{KR_pot_q}
  V_{adj}^B (q) \, = \, \frac{2 \, \lambda}{N_c^2} \,
  \int\limits_0^{1} dz \, {\cal B}_{0z} (q,z)
\end{align}
with $z$ and $q$ now taken in the units of $z_h$.

%%%%%%%%%%%%%%%%%%%%%%%%%%%%%%%%%%%%%%%%%%%

\subsubsection{Solution of the 2-form EOM}

The series solution of \eq{KR_EOM4} is slightly more involved than
that for the dilaton and $t^0$ and is given by
\begin{align}
  \label{KR_series}
  {\cal B}_{0z} \, = \, \sum_{n=0}^\infty \, d_n \, z^{2 \, n +1} +
  \sum_{n=1}^\infty l_n \, z^{2 \, n +1} \, \ln z + C^B (q^2) \,
  \sum_{n=0}^\infty \, e_n \, z^{2 \, n +3}
\end{align}
with the recursion relations
\begin{subequations}
\begin{align}
  \label{d_rec}
  d_n \, = \, \frac{q^2}{4 \, (n-1) \, (n +3)} \, d_{n-1} +
  \frac{n+1}{n+3} \, d_{n-2} - \frac{n+1}{(n-1) \, (n+3)} \, l_n +
  \frac{n}{(n-1) \, (n+3)} \, l_{n-2}, \notag \\ d_0 = - \frac{1}{12},
  \ \ d_1 = 0,
\end{align}
\begin{align}
  \label{l_rec}
  l_n \, = \, \frac{q^2}{4 \, (n-1) \, (n+2)} \, l_{n-1} +
  \frac{n+1}{n+3} \, l_{n-2}, \ \ \ \ \ l_0 = 0, \ \ l_1 = -
  \frac{q^2}{96},
\end{align}
\begin{align}
  \label{e_rec}
  e_n \, = \, \frac{q^2}{4 \, n \, (n+4)} \, e_{n-1} + \frac{n+2}{n+4}
  \, e_{n-2}, \ \ \ \ \ e_{-1} = 0, \ \ e_0 = 1.
\end{align}
\end{subequations}
Neumann boundary conditions for the partial sums of the series give
\begin{align}
  \label{CKR}
  C^B (q^2, N) \, = \, - \frac{\sum_{n=0}^N \, d_n \, (2 \, n +1) +
    \sum_{n=1}^N \, l_n}{\sum_{n=0}^N \, e_n \, (2 \, n +3)}.
\end{align}

%%%%%%%%%%%%%%%%%%%%%%%%%%%%%%%%%%%%%%%%%%%%%%%%%%%%%%%%%%%
\FIGURE{\includegraphics[width=12cm]{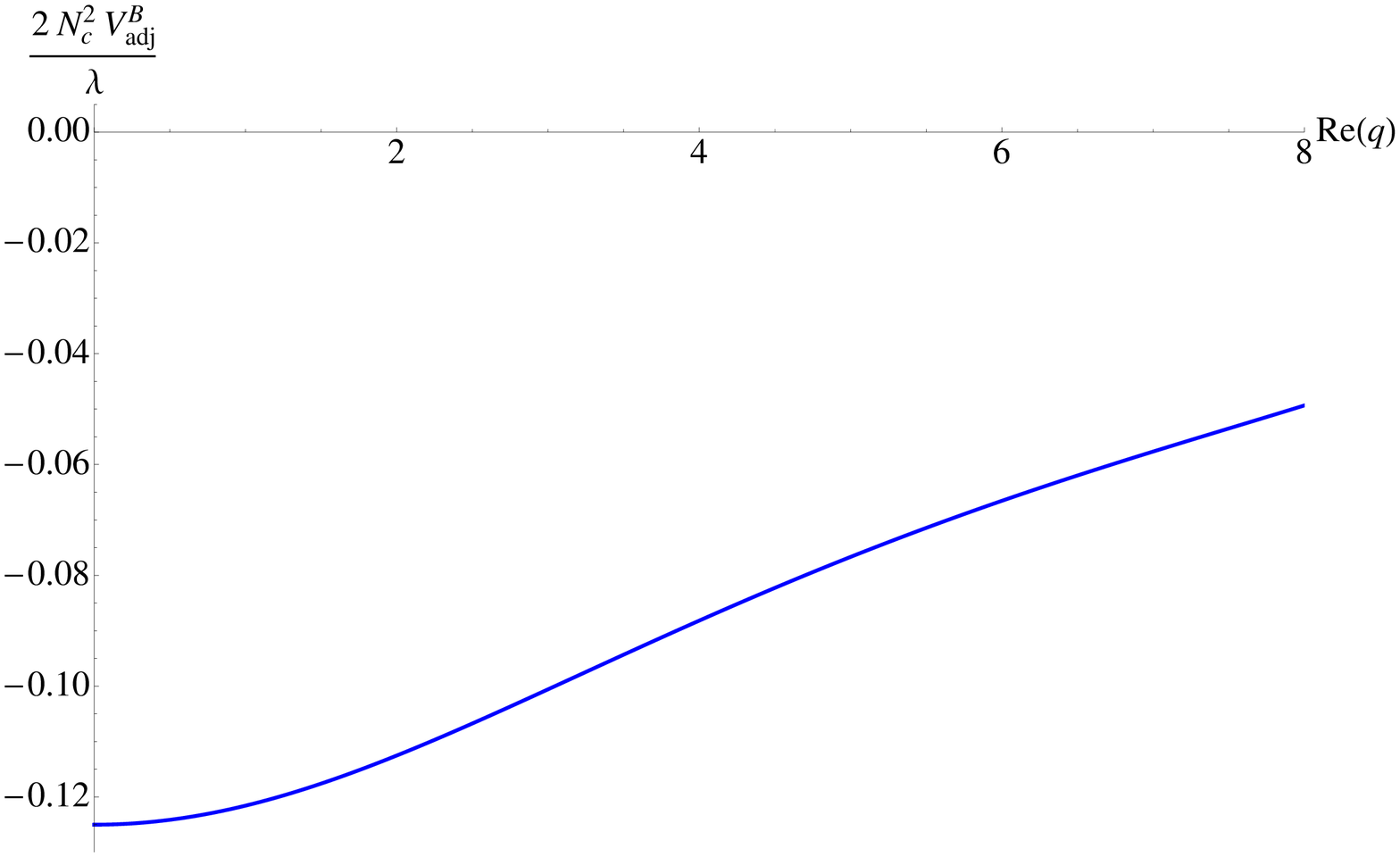}
  \caption{The contribution to the adjoint finite-$T$ $Q\bar Q$ 
    potential in momentum space due to the exchange of the 2-form
    field between the string world-sheets, plotted along the real-$q$
    axis (Im~$q=0$) in units of $\lambda/(2 \, N_c^2)$. The plot is for $N= 2 \times 10^4$ iterations.}
  \label{KRre}
}
%%%%%%%%%%%%%%%%%%%%%%%%%%%%%%%%%%%%%%%%%%%%%%%%%%%%%%%%%%%

Plugging \eq{KR_series} into \eq{KR_pot_q} we obtain a series
representation of the contribution of the 2-form field to the heavy
quark potential in momentum space
\begin{align}
  \label{VKR_series}
  V_{adj}^B (q) \, = \, \frac{2 \, \lambda}{N_c^2} \, \left[
    \sum_{n=0}^\infty \, \frac{d_n}{2 \, n +2} - \sum_{n=1}^\infty
    \frac{l_n}{4 \, (n +1)^2} + C^B (q^2) \, \sum_{n=0}^\infty \,
    \frac{e_n}{2 \, n +4} \right].
\end{align}
Our evaluation of this potential by numerically summing up the series
is plotted in \fig{KRre} in units of $\lambda/(2\, N_c^2)$. Just like
the dilaton and $t^0$ contributions, the 2-form potential in
\fig{KRre} appears to be {\sl attractive}.

%%%%%%%%%%%%%%%%%%%%%%%%%%%%%%%%%%%%%%%%%%%%%%%%%

\subsubsection{Asymptotics of the 2-form contribution}

As before the 2-form contribution $V_{adj}^B (q)$ only has poles along
the imaginary-$q$ axis. The plot of $V_{adj}^B (q)$ along the positive
Im~$q$-axis is shown in \fig{KRim}.

%%%%%%%%%%%%%%%%%%%%%%%%%%%%%%%%%%%%%%%%%%%%%%%%%%%%%%%%%%%
\FIGURE{\includegraphics[width=12cm]{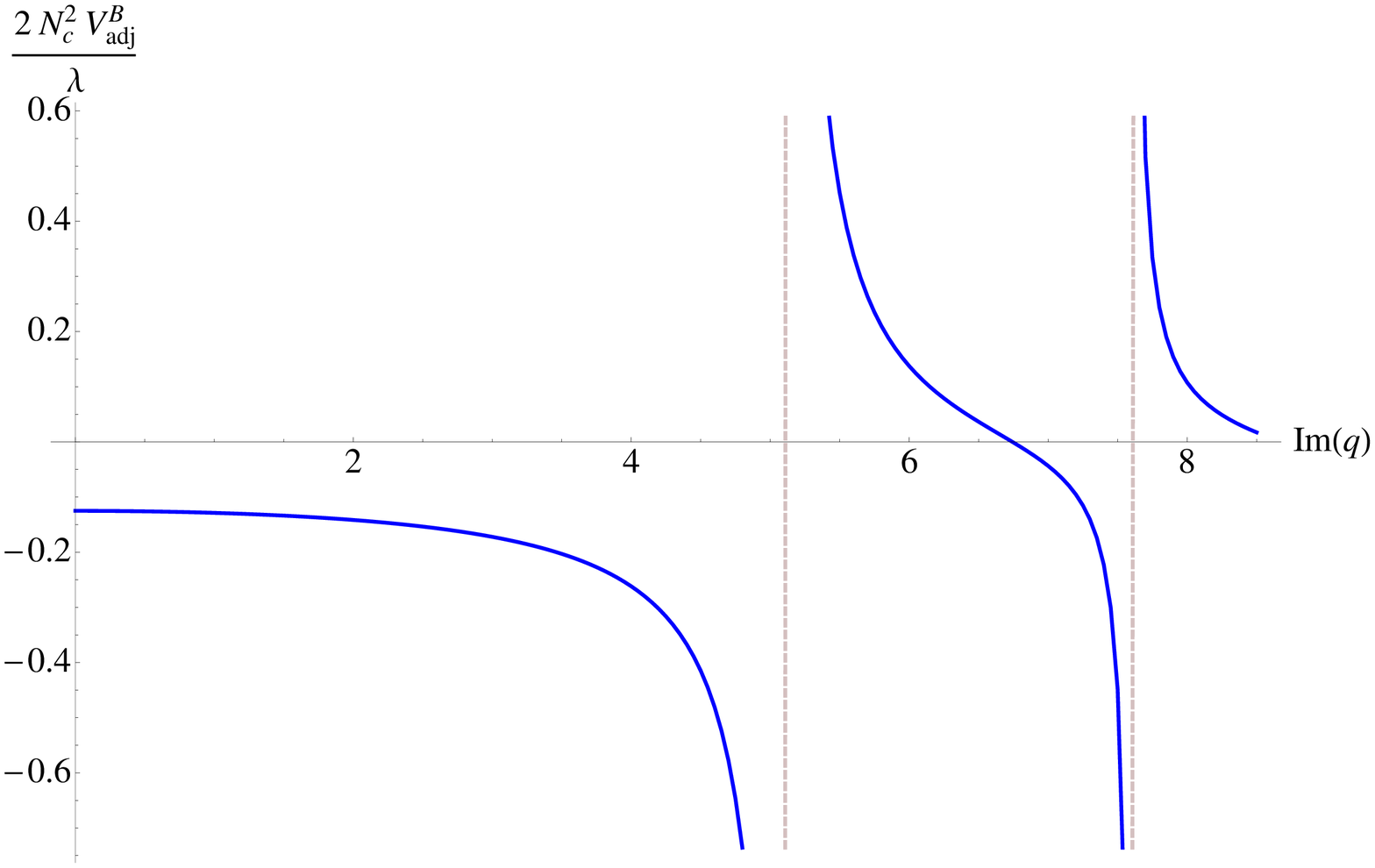}
  \caption{The contribution to the adjoint finite-$T$ $Q\bar Q$ 
    potential in momentum space due to the exchange of the 2-form
    field between the string world-sheets, plotted along the
    imaginary-$q$ axis (Re~$q=0$) in units of $\lambda/(2 \, N_c^2)$. The plot is for $N=10^4$ iterations.}
  \label{KRim}
}
%%%%%%%%%%%%%%%%%%%%%%%%%%%%%%%%%%%%%%%%%%%%%%%%%%%%%%%%%%%

The leading pole of $V_{adj}^B (q)$ is found to be at $m_1^B
= 5.1086 \pm 0.0001$ (for $N= 8 \times 10^5$ iterations), in agreement with \cite{Brower:2000rp}.  The
residue of this pole is $i \, (0.051 \pm 0.003)$, such that the
large-$r\, T$ asymptotics of the 2-form contribution to the
coordinate-space potential is
\begin{align}
  \label{large_B}
  V_{adj}^B (r) \bigg|_{r \, T \gg 1} \, \approx \, - \frac{2 \,
    \lambda}{N_c^2} \, \frac{0.26 \pm 0.02}{2 \, \pi \, r} \, e^{-
    5.1086 \, \pi \, r \, T}.
\end{align}
Again this is an attractive potential. 

At small-$r\, T$/large-$q$ the asymptotic solution of \eq{KR_EOM4} is
\begin{align}
  \label{KRsol_large-q}
  {\cal B}_{0z} \big|_{q \, z \gg 1} \, \approx \, - \frac{1}{q^2 \, z},
\end{align}
which, when used in Eqs.~\peq{KR_pot_q} and \peq{potF} yields
\begin{align}
  \label{KRpot_small-r}
  V_{adj}^B (r) \bigg|_{r \, T \ll 1} \, \approx \, - \frac{2 \,
    \lambda}{N_c^2} \, \frac{1}{4 \, \pi \, r} \, \ln \frac{1}{r \,
    T},
\end{align}
which is also attractive.

%%%%%%%%%%%%%%%%%%%%%%%%%%%%%%%%%%%%%%%%%%%%%%%%%%%%%%%%%%%%%%%%%%%%%%%%%%%%%%%%%%%

\subsection{The Graviton}

%%%%%%%%%%%%%%%%%%%%%%%%%%%%%%%%%%%%%%%%%%%

\subsubsection{The potential and Einstein equations}

We want to find quadratic fluctuations in the metric, $g_{MN} +
h_{MN}$, induced in the presence of a straight string stretching from
the UV boundary all the way down to the horizon.  The most general
form of the static metric with the $SO(3)$ rotational symmetry takes
the form
\begin{align}\label{metricgeneral}
  ds^2 = \frac{f(z)}{z^2}\left[1 + h(r,z)\right]d\tau^2 + \frac{1 +
    A(r,z)}{z^2}\left[dr^2 + r^2 d\Omega_2^2\right] +
  \frac{1+B(r,z)}{z^2 \, f(z)} \, dz^2 \ .
\end{align}
It can be shown that there are consistent (non-singular)
diffeomorphism transformations that can lead to the above metric.

Since the coupling of the string world-sheet to graviton is
\begin{align}\label{gravstring}
  S_{h-str} = \frac{1}{4\pi\alpha'} \, \int d^2\sigma~\sqrt{\gamma} \,
  \gamma^{ab} \, h_{MN} \, \partial_aX^M \, \partial_bX^N
\end{align}
it follows that the 5-dimensional energy-momentum (EM) tensor of the
string at ${\vec X}_2 =0$ in the bulk is
\begin{align}
  J^{MN} = \frac{1}{2\pi\alpha'} \, \frac{\sqrt{\gamma}}{\sqrt{g}} \,
  \gamma^{ab} \, \delta^{(3)}(\vec{x}) \, \partial_aX^M \,
  \partial_bX^N
\end{align}
(see \cite{Friess:2006fk,Lin:2007pv,Chesler:2007sv} for similar
calculations). The non-vanishing components of the EM tensor are
\begin{align}
  J^{0}_0 = J^z_z = \frac{z^3}{2\, \pi\, \alpha'} \,
  \delta^{(3)}(\vec{x}) \ .
\end{align}
Therefore, in the presence of the string source, the Einstein
equations can be written as
\begin{align}\label{ein}
  \cG^M_N \, = \, \left(6 + \kappa_5^2 \, J^0_0 \right) \,
  \delta^{M}_{N} \ ,
\end{align}
where $\cG^M_N$ is the Einstein tensor, $6$ is coming from the
cosmological constant term, and $\kappa_5^2 = 4 \, \pi^2 / N_c^2$. As
usual we will work in units where $z_h=1$ and $R_{\rm AdS} = 1$.
Linearized Einstein equations \peq{ein} for the metric
\peq{metricgeneral} are presented and simplified in Appendix \ref{A},
using momentum-space metric components
\begin{align}
  \label{FTG}
  \bar{A} (r, z) \, = \, \int \frac{d^3 q}{(2 \, \pi)^3} & \ e^{i \,
    {\vec q} \cdot {\vec r}} \ {\cal A} (q, z), \ \ \ \bar{B} (r, z)
  \, = \, \int \frac{d^3 q}{(2 \, \pi)^3} \,
  e^{i \, {\vec q} \cdot {\vec r}} \ {\cal B} (q, z), \notag \\
  & \bar{h} (r, z) \, = \, \int \frac{d^3 q}{(2 \, \pi)^3} \, e^{i \,
    {\vec q} \cdot {\vec r}} \ \tgh (q, z)
\end{align}
for the rescaled fields
\begin{align}
  \label{Gresc}
  \bar{A} ({\vec r}, z) \, \equiv \, \frac{2\pi\alpha'}{\kappa^2_5} \,
  A, \ \ \ \bar{B} ({\vec r}, z) \, \equiv \,
  \frac{2\pi\alpha'}{\kappa^2_5} \, B, \ \ \ \bar{h} ({\vec r}, z) \,
  \equiv \, \frac{2\pi\alpha'}{\kappa^2_5} \, h.
\end{align}
As argued in Appendix \ref{A}, solving Einstein equations with
additional gauge freedom left in the metric \peq{metricgeneral} allows
one to set
\begin{align}
  \label{ABh}
  \bar{A} + \bar{B} + \bar{h} \, = \, 0.
\end{align}
After lengthy but straightforward calculations presented in Appendix
\ref{A}, the equation for $\bar{A}$ (at the leading order in
perturbations) can be brought to the following form:
\begin{align}\label{GR_EOM}
  z \, (1 - z^4) \, (3 - 3 z^4 - q^2 \, z^2) \, \cA_{z \, z} & -
  \left[ 3 \, (1 - z^4)^2 - q^2 \, z^2 \, (3 + z^4) \right] \, \cA_z
  \notag \\ & - q^2 \, z \, \left[1 - q^2 \, z^2 + 3 \, z^4 \right] \,
  \cA \, = \, \frac{4}{3} \, q^2 z^4 \ .
\end{align}
Again the boundary conditions are $\cA(q,z=0) = 0$ and $\cA (q,z=1)$
should be finite.

Similar to the previous cases, the full graviton action evaluated on
the classical solution of \eq{GR_EOM} is
\begin{align}\label{gravstringSol}
  \bar{S}_{h} = \frac{1}{4\pi\alpha'}\int\limits_{(1)}
  d^2\sigma~\sqrt{\gamma}\gamma^{ab} h_{MN}\partial_aX^M \partial_bX^N
  = -\frac{\beta \, \lambda}{2 \, N^2_c} \, \int\limits^{1}_0 \,
  \frac{dz}{z^2} \, \bar{A} (r,z) \ ,
\end{align}
such that the graviton contribution to the adjoint heavy quark
potential is
\begin{align}
  \label{V_GR_r}
  V_{adj}^G (r) \, = \, - \frac{\lambda}{2 \, N^2_c} \,
  \int\limits^{1}_0 \, \frac{dz}{z^2} \, \bar{A} (r,z)
\end{align}
in coordinate space, translating into
\begin{align}
  \label{V_GR_q}
  V_{adj}^G (q) \, = \, - \frac{\lambda}{2 \, N^2_c} \,
  \int\limits^{1}_0 \, \frac{dz}{z^2} \, \cA (q,z)
\end{align}
in momentum space.

%%%%%%%%%%%%%%%%%%%%%%%%%%%%%%%%%%%%%%%%%%%

\subsubsection{Solution of Einstein equations}

The series solution of \eq{GR_EOM} is
\begin{align}
  \label{GR_series}
  \cA \, = \, \sum_{n=0}^\infty \, p_n \, z^{2 \, n + 5} + C^G (q^2)
  \, \sum_{n=0}^\infty \, s_n \, z^{2 \, n + 2}
\end{align}
with the recursion relations
\begin{subequations}\label{GR_rec}
  \begin{align}
    \label{p_rec}
    p_n \, = \, & \frac{2 \, (2 \, n^2 + 2 \, n -1)}{3 \, (2 \, n +3)
      \, (2 \, n + 5)} \, q^2 \, p_{n-1} - \frac{6 - 24 \, n^2 +
      q^4}{3 \, (2 \, n +3) \, (2 \, n + 5)} \, p_{n-2} \notag \\ & -
    \frac{2 \, (2 \, n^2 - 2 \, n -1)}{3 \, (2 \, n +3) \, (2 \, n +
      5)} \, q^2 \, p_{n-3} - \frac{(2 \, n - 3) \, (2 \, n - 5)}{(2
      \, n +3) \, (2 \, n + 5)} \, p_{n-4}, \notag \\ & p_0 = \frac{4
      \, q^2}{135}, \ p_{-1}= p_{-2} = p_{-3} = 0, \\
    s_n \, = \, & \frac{4 \, n^2 - 8 \, n +1}{12 \, n \, (n + 1)} \,
    q^2 \, s_{n-1} + \frac{48 - 72 \, n + 24 \, n^2 - q^4}{12 \, n \,
      (n + 1)} \, s_{n-2} \notag \\ & - \frac{13 - 16 \, n + 4 \,
      n^2}{12 \, n \, (n + 1)} \, q^2 \, s_{n-3} - \frac{(n - 3) \, (n
      - 4)}{n \, (n + 1)} \, s_{n-4}, \notag \\ & s_0 = 1, \ s_{-1}=
    s_{-2} = s_{-3} = 0.
    \label{s_rec}
  \end{align}
\end{subequations}
Neumann boundary conditions imposed on partial sums of the series
\peq{GR_series} yield
\begin{align}
  \label{CGR}
  C^G (q^2, N) \, = \, - \frac{\sum_{n=0}^N \, p_n \, (2 \, n +
    5)}{\sum_{n=0}^N \, s_n \, (2 \, n + 2)}.
\end{align}

%%%%%%%%%%%%%%%%%%%%%%%%%%%%%%%%%%%%%%%%%%%%%%%%%%%%%%%%%%%
\FIGURE{\includegraphics[width=12cm]{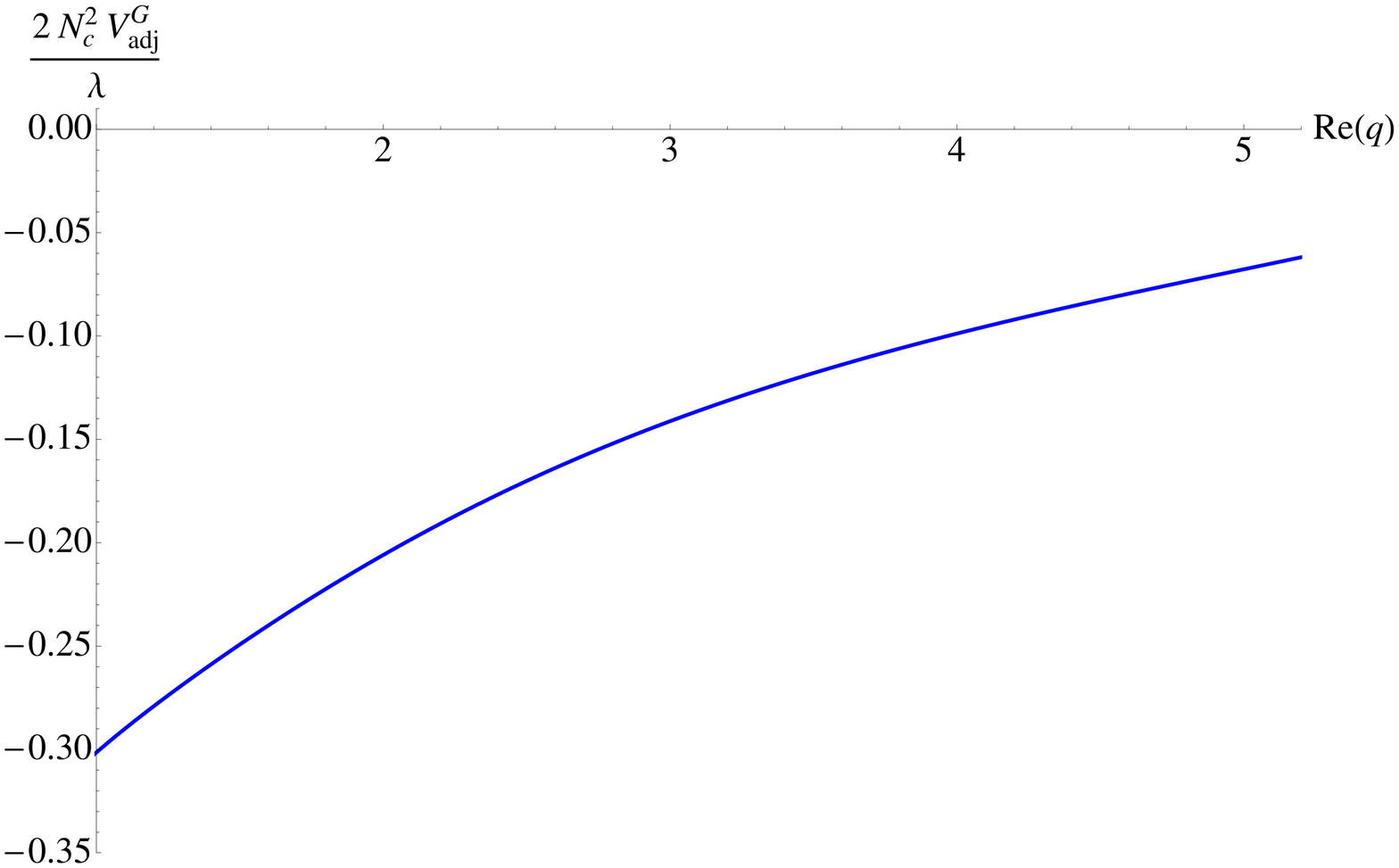}
  \caption{The contribution to the adjoint finite-$T$ $Q\bar Q$ 
    potential in momentum space due to the exchange of the graviton
    field between the string world-sheets, plotted along the real-$q$
    axis (Im~$q=0$) in units of $\lambda/(2 \, N_c^2)$. The plot is for $N=10^3$ iterations.}
  \label{GRre}
}
%%%%%%%%%%%%%%%%%%%%%%%%%%%%%%%%%%%%%%%%%%%%%%%%%%%%%%%%%%%

Finally, the series representation for the potential is obtained by
using \eq{GR_series} in \eq{V_GR_q}, which results in
\begin{align}
  \label{VGR_series_q}
  V_{adj}^G (q) \, = \, - \frac{\lambda}{2 \, N^2_c} \,
  \left[ \sum_{n=0}^\infty \, \frac{p_n}{2 \, (n+2)} + C^G (q^2) \,
    \sum_{n=0}^\infty \, \frac{s_n}{2 \, n + 1} \right].
\end{align}
The numerical evaluation of the series is plotted in \fig{GRre}. The
graviton naturally gives an attractive potential. 

The $q$-range of the plot in \fig{GRre} is limited by the convergence
of the series in \eq{VGR_series_q} both from above and from below,
unlike the previous supergravity fields for which the convergence
problems only appeared at large-$q$. At small (but non-zero) real $q$
the numerical evaluation of the partial sums of the series in
\eq{VGR_series_q} appears to diverge, though the singularity moves
towards $q=0$ as one increases the number of terms in the sums. One
may then suspect a non-analyticity at $q=0$ in $V_{adj}^G (q)$. On the
other hand one has to remember that in arriving at \eq{GR_EOM} we have
made a substitution \peq{BA}, which is potentially singular at $q=0$:
the peculiar behavior of the numerics may be attributed to this
potentially dangerous operation. To test whether this convergence
issue is an artifact of the numerics or a genuine singularity of
$V_{adj}^G (q)$ at $q=0$ one may search for the solution of
\eq{GR_EOM} as a power-series in $q^2$ with the coefficient being some
functions of $z$. A straightforward calculation yields
\begin{align}
  \label{Aexp_q}
  \cA (q, z) \, = \, \frac{z^2}{3} + O (q^2),
\end{align}
where, in order to fix the coefficient of the leading term one has to
require finiteness of $\cA (q, z=1)$ at the order-$q^2$. Substituting
\eq{Aexp_q} into \eq{V_GR_q} gives
\begin{align}
  \label{VGR0}
  V_{adj}^G (q=0) \, = \, - \frac{\lambda}{6 \, N^2_c}.
\end{align}
We see that the graviton contribution to the heavy quark potential is
finite at $q=0$. Moreover, the $q^2$ series in \eq{Aexp_q} can be
easily continued, demonstrating that $V_{adj}^G (q)$ is analytic at
$q=0$. Therefore the divergence described above is a numerical
artifact and not a real physical divergence.

%%%%%%%%%%%%%%%%%%%%%%%%%%%%%%%%%%%%%%%%%%%%%%%%%

\subsubsection{Asymptotics of the graviton contribution}

The graviton contribution $V_{adj}^G (q)$ only has poles
along the imaginary-$q$ axis. The plot of $V_{adj}^G (q)$
along this axis is shown in \fig{GRim}.

%%%%%%%%%%%%%%%%%%%%%%%%%%%%%%%%%%%%%%%%%%%%%%%%%%%%%%%%%%%
\FIGURE{\includegraphics[width=12cm]{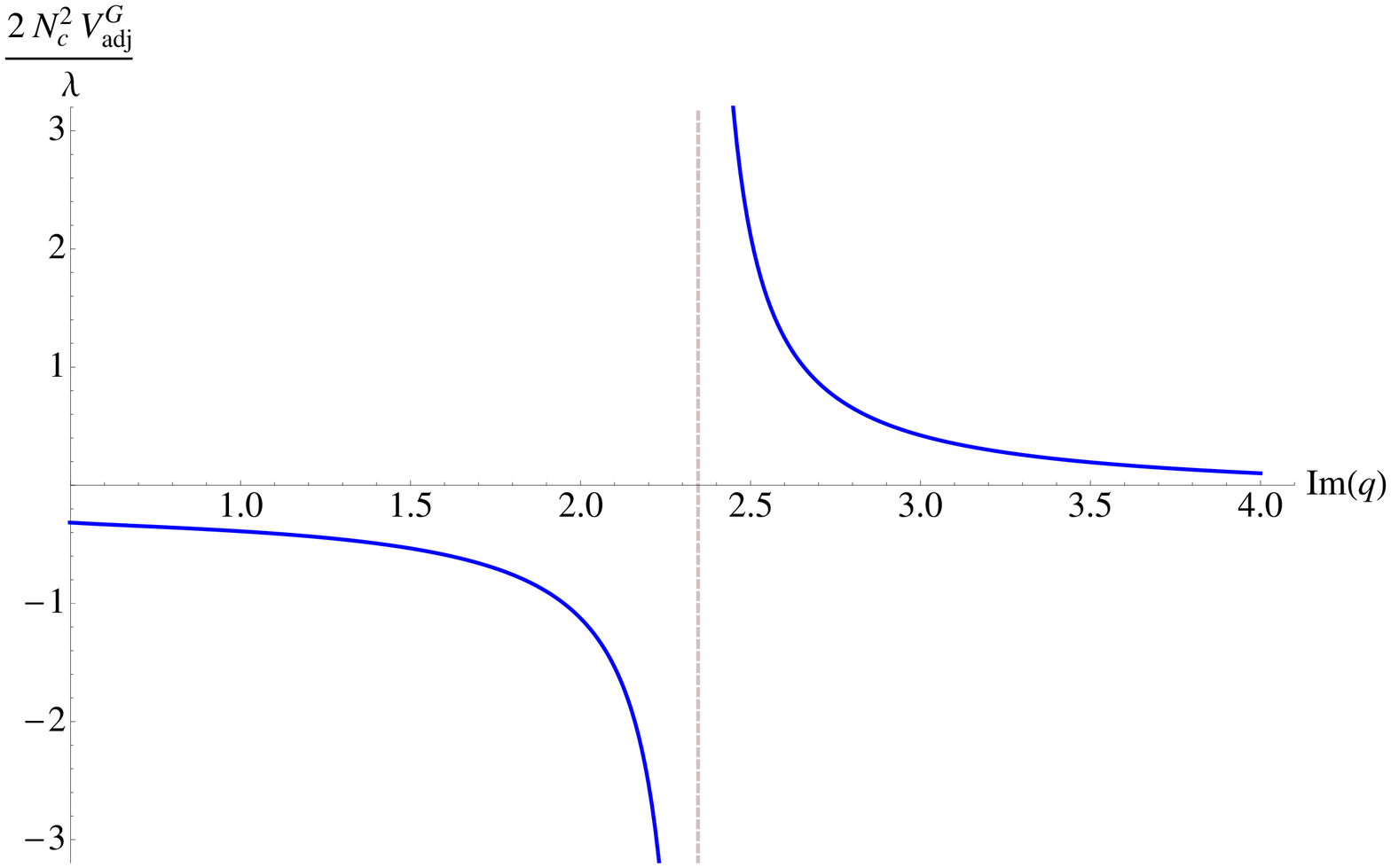}
  \caption{The contribution to the adjoint finite-$T$ $Q\bar Q$ 
    potential in momentum space due to the exchange of the graviton
    field between the string world-sheets, plotted along the
    imaginary-$q$ axis (Re~$q=0$) in units of $\lambda/(2 \, N_c^2)$. The plot is for $N=10^3$ iterations.}
  \label{GRim}
}
%%%%%%%%%%%%%%%%%%%%%%%%%%%%%%%%%%%%%%%%%%%%%%%%%%%%%%%%%%%

The leading pole of $V_{adj}^G (q)$ is at $m_1^G = 2.33 \pm
0.01$ (for $N= 3000$ iterations), also in agreement with \cite{Csaki:1998qr,Brower:2000rp},
though with significantly lower precision, due to the complicated
structure of the recurrence relations \peq{GR_rec} for the series
coefficients.  The residue of the leading pole is $i \, (0.37 \pm
0.01)$, giving the following large-$r\, T$ asymptotics of the graviton
contribution to the coordinate-space potential
\begin{align}
  \label{large_G}
  V_{adj}^G (r) \bigg|_{r \, T \gg 1} \, \approx \, - \frac{\lambda}{2
    \, N_c^2} \, \frac{0.86 \pm 0.03}{2 \, \pi \, r} \, e^{- 2.33 \,
    \pi \, r \, T}.
\end{align}
This is also an attractive-potential contribution.

At small-$r\, T$/large-$q$ the asymptotic solution of \eq{GR_EOM} is
\begin{align}
  \label{GRsol_large-q}
  {\cal A} \big|_{q \, z \gg 1} \, \approx \, \frac{4}{3} \,
  \frac{z}{q^2},
\end{align}
which, when used in \eq{V_GR_q} yields
\begin{align}
  \label{GRpot_small-r}
  V_{adj}^G (r) \bigg|_{r \, T \ll 1} \, \approx \, - \frac{2}{3} \,
  \frac{\lambda}{N_c^2} \, \frac{1}{4 \, \pi \, r} \, \ln \frac{1}{r
    \, T},
\end{align}
which is, yet again, an attractive potential.

%%%%%%%%%%%%%%%%%%%%%%%%%%%%%%%%%%%%%%%%%%%%%%%%%%%%%%%%%%%%%%%%%%%%%%%%%%%%%%%%%%%

\subsection{The Net Result}

The net contribution of all supergravity fields to the $Q\bar Q$
potential at the order-$\lambda/N_c^2$ is given by adding together
Eqs.~\peq{pot_dil_q}, \peq{pot_t0_q}, \peq{KR_pot_q}, and
\peq{V_GR_q}, which yields
\begin{align}
  \label{netVq}
  V_{adj}^{Q\bar Q} (q) \, & = \, V_{adj}^\phi (q) +
  V_{adj}^{t^0} (q) + V_{adj}^B (q) + {\tilde
    V}_{adj}^G (q) \notag \\ & = \, \frac{\lambda}{2 \, N_c^2} \,
  \int\limits_0^1 \, \frac{d z}{z^2} \, \left[ \varphi (q, z) +
    \frac{5}{3} \, \tgt^0 (q, z) + 4 \, z^2 \, {\cal B}_{0z}
    (q, z) - \cA (q, z) \right].
\end{align}
The corresponding coordinate-space potential $V_{adj}^{Q\bar Q} (r)$,
\begin{align}
  \label{netVr}
  V_{adj}^{Q\bar Q} (r) \, = \, V_{adj}^\phi (r) + V_{adj}^{t^0} (r) +
  V_{adj}^B (r) + V_{adj}^G (r),
\end{align}
can be obtained from \eq{netVq} using \eq{potF}.  The total potential
$V_{adj} (q)$ is plotted in \fig{totalpot} (lower line) in
momentum space (along with the quark-quark potential to be discussed
later in Sec.  \ref{QQ}).  Since all the four contributions to this
potential were attractive, the resulting potential is indeed
attractive. This is in contrast to the adjoint potential in
perturbation
%%%%%%%%%%%%%%%%%%%%%%%%%%%%%%%%%%%%%%%%%%%%%%%%%%%%%%%%%%%
\FIGURE{\includegraphics[width=14cm]{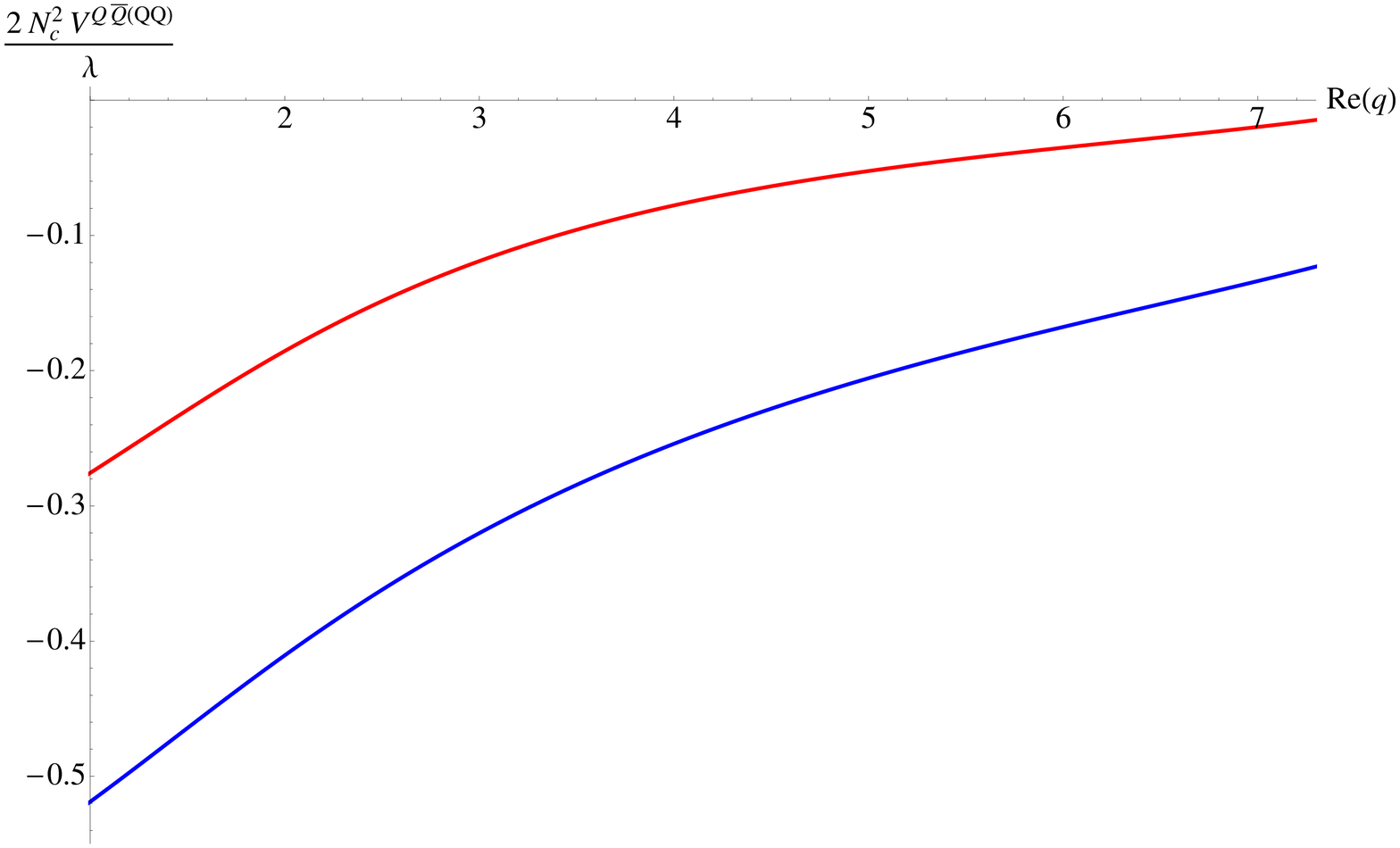}
  \caption{The net adjoint finite-$T$ $Q\bar Q$ potential (lower curve) and the 
    net QQ potential (upper curve) in momentum space due to the
    exchanges of all supergravity fields between the string
    world-sheets, plotted along the real-$q$ axis (Im~$q=0$) in units
    of $\lambda/(2 \, N_c^2)$.}
  \label{totalpot}
}
%%%%%%%%%%%%%%%%%%%%%%%%%%%%%%%%%%%%%%%%%%%%%%%%%%%%%%%%%%%
theory, which was found to be repulsive (see e.g.
\cite{Nadkarni:1986cz}). As observed before, the adjoint potential (or
the adjoint contribution to the unified total potential, as discussed
in Sec. \ref{One}) is of order-$\lambda/N_c^2$ both at weak and at
strong couplings. We therefore conjecture that the difference between
the strong- and weak-coupling regimes is in the sign of the
interaction, making the repulsive weak-coupling potential attractive
at strong coupling.

The large-$r \, T$ asymptotics of the net potential $V_{adj}^{Q\bar Q}
(r)$ is given by the mode with the lightest screening mass, which, as
follows from the above calculations (and from earlier works
\cite{Csaki:1998qr,Brower:2000rp}) is the graviton, such that
\begin{align}
  \label{net_large_r}
  V_{adj}^{Q\bar Q} (r) \bigg|_{r \, T \gg 1} \, \approx \, V_{adj}^G
  (r) \bigg|_{r \, T \gg 1} \, \approx \, - \frac{\lambda}{2 \, N_c^2}
  \, \frac{0.86 \pm 0.03}{2 \, \pi \, r} \, e^{- 2.33 \, \pi \, r \,
    T},
\end{align}
with the exponent in agreement with \cite{Bak:2007fk}. As we discussed
above, if higher KK modes were allowed, than the large-$r \, T$
scaling would be dominated by the $s^2$ field, and would be given by
\eq{pot_s2_rlarge}.

At small-$r\, T$, adding the contributions from
Eqs.~\peq{Vdil-small-r}, \peq{Vt0-small-r}, \peq{KRpot_small-r}, and
\peq{GRpot_small-r} yields
\begin{align}
  \label{net-small-r}
  V_{adj}^{Q\bar Q} (r) \bigg|_{r \, T \ll 1} \, \approx \, - 4 \,
  \frac{\lambda}{N_c^2} \, \frac{1}{4 \, \pi \, r} \, \ln \frac{1}{r
    \, T}.
\end{align}

Let us now return to the question of what happens to the potential
$V_{adj}^{Q\bar Q} (r)$ as $r \, T \rightarrow 0$. As follows from,
say, \eq{netVq}, the potential is obtained by integrating the
contribution of various supergravity fields over the string
world-sheet ($dz/z^2$). Our setup of static straight stings exchanging
supergravity fields is valid only if the fields are weak. In
coordinate space in the small-$r\, T$ limit the fields are
parametrically of the order
\begin{align}
  \frac{\lambda}{N_c^2} \, \frac{z}{r}
\end{align}
as follows from Fourier-transforming e.g. \eq{dil_large-q} (we include
the coupling to the other string as well). The applicability region of
the weak-field approximation is then defined by
\begin{align}
  \frac{\lambda}{N_c^2} \, \frac{z}{r} \ll 1, 
\end{align}
which means that
\begin{align}
  \label{z_lim}
  z \ll r \, \frac{N_c^2}{\lambda}.
\end{align}
Therefore, assuming that corrections to our setup, coming presumably
from string fluctuations, would regularize the divergence in the $z$
integral at large-$z$ (see e.g.  \cite{Erickson:1999qv} for similar
phenomena), and putting $z_h=1/(\pi \, T)$ back into the expression
explicitly, we see that the $z$-integral in \eq{netVq} is cut off by
min~$\left\{ z_h, r \, N_c^2/ \lambda \right\}$ in the IR, such that
\begin{align}
  \label{net-small-r2}
  V_{adj}^{Q\bar Q} (r) \bigg|_{r \, T \ll 1} \, \approx \, - 4 \,
  \frac{\lambda}{N_c^2} \, \frac{1}{4 \, \pi \, r} \, \ln \left(
    \frac{1}{r} \, \text{min} \, \left\{ z_h, r \,
      \frac{N_c^2}{\lambda} \right\} \right),
\end{align}
again with the logarithmic accuracy. Therefore we expect
\begin{align}
  \label{net-small-r3}
  V_{adj}^{Q\bar Q} (r) \bigg|_{r \, T \ll \lambda/N_c^2} \, \approx
  \, - 4 \, \frac{\lambda}{N_c^2} \, \frac{1}{4 \, \pi \, r} \, \ln
  \left( \frac{N_c^2}{\lambda} \right),
\end{align}
which is finite in the $r\, T \rightarrow 0$ limit. 

%%%%%%%%%%%%%%%%%%%%%%%%%%%%%%%%%%%%%%%%%%%%%%%%%%%%%%%%%%%%%%%%%%%%%%%%

%%%%%%%%%%%%%%%%%%%%%%%%%%%%%%%%%%%%%%%%%%%%%%%%%%%%%%%%%%%%%%%%%%%%%%%%%%%%%%%
%%%%%%%%%%%%%%%%%%%%%%%%%%%%%%%%%%%%%%%%%%%%%%%%%%%%%%%%%%%%%%%%%%%%%%%%%%%%%%%

\section{The $Q Q$ potential}
\label{QQ}

We can use the results of the above calculation to find the potential
between two heavy quarks immersed in the finite-$T$ ${\cal N} =4$ SYM
medium.  While the decomposition into the two potentials corresponding
to color group representations $\bar{N_c}$ and $N_c^2 - N_c$ is
possible along the lines of \eq{decomp}, as we will see momentarily
both potentials obtained this way would correspond to the same string
configuration in AdS space, and, therefore, would be equal to each
other. Hence we define the $QQ$ potential simply as
\begin{align}
  \label{QQ_pot_def}
  e^{- \beta \, V^{QQ} (r)} \, = \, \frac{1}{N_c^2} \, \left\langle
    \text{Tr} \, L (0) \ \text{Tr} \, L ({\vec r}) \right\rangle_c
\end{align}
in analogy to \eq{one_pot}. 

%%%%%%%%%%%%%%%%%%%%%%%%%%%%%%%%%%%%%%%%%%%%%%%%%%%%%%%%%%%
\FIGURE{\includegraphics[width=12cm]{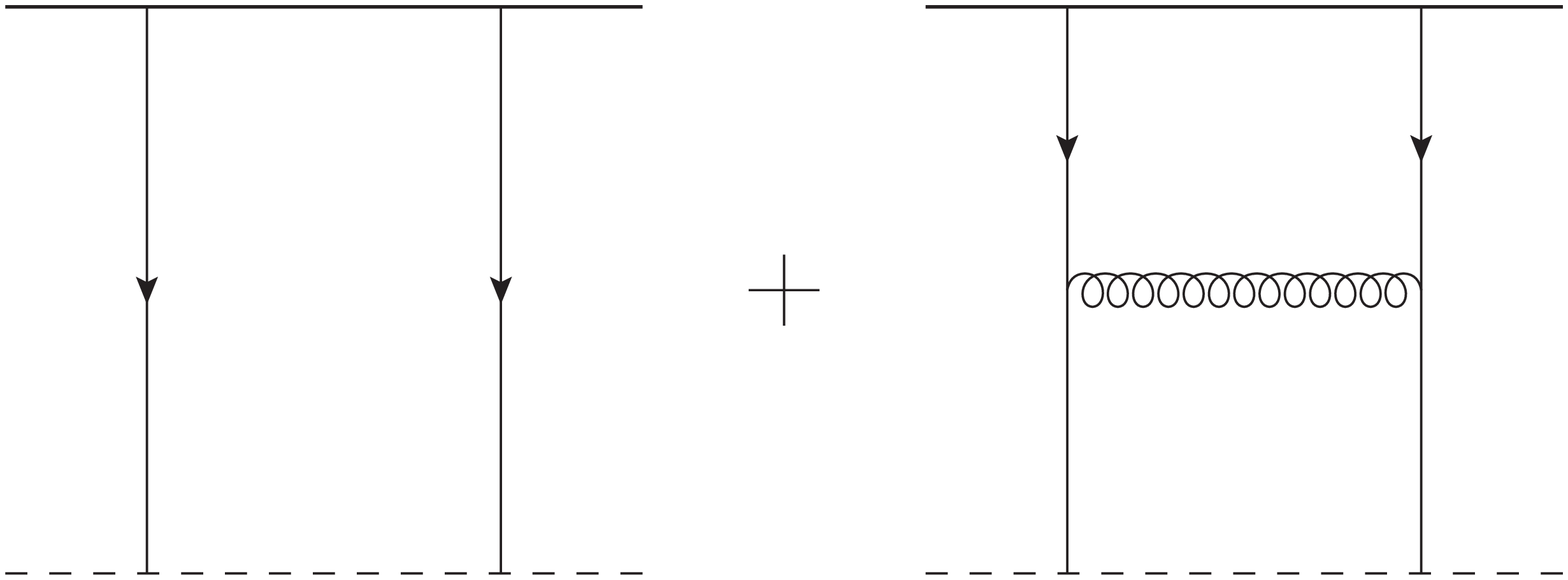}
  \caption{The quark--quark potential at finite-$T$ at the order-$\lambda/N_c^2$.}
  \label{qq_pot}
}
%%%%%%%%%%%%%%%%%%%%%%%%%%%%%%%%%%%%%%%%%%%%%%%%%%%%%%%%%%%

To calculate $V^{QQ} (r)$ using AdS/CFT we notice that now both
strings attached to the quarks have to have the same orientation.
Therefore the hanging string configuration from the left panel of
\fig{strings} is now impossible, and the potential is given by the
straight-strings configuration shown in \fig{qq_pot}. Just as for the
$Q\bar Q$ system, the left panel in \fig{qq_pot} gives zero
contribution to the potential after renormalization and subtraction of
self-interactions. We are left with the right panel in \fig{qq_pot}.
Note that now both strings are oriented in the same direction, and can
be parametrized as (cf. \eq{strings_param})
\begin{align}
  \label{strings_param_qq}
  X_1^M = (\tau, {\vec r}, z=\sigma), \ \ \ \ \ X_2^M = (\tau, {\vec
    0}, z = \sigma).
\end{align}
The effect of this difference is only felt by the 2-form field
coupling to the string $X_2$, which changes its sign, as follows from
\eq{KRaction}. Hence the action of the configuration in the right
panel of \fig{qq_pot} differs from that in \fig{exchange} by the sign
of the 2-form contribution. We therefore write for the momentum-space
$QQ$ potential
\begin{align}
  \label{netVqq}
  V^{QQ} (q) \, & = \, V_{adj}^\phi (q) +
  V_{adj}^{t^0} (q) - V_{adj}^B (q) + {\tilde
    V}_{adj}^G (q) \notag \\ & = \, \frac{\lambda}{2 \, N_c^2} \,
  \int\limits_0^1 \, \frac{d z}{z^2} \, \left[ \varphi (q, z) +
    \frac{5}{3} \, \tgt^0 (q, z) - 4 \, z^2 \, {\cal B}_{0z}
    (q, z) - \cA (q, z) \right]
\end{align}
and for the coordinate-space potential
\begin{align}
  \label{netVr_qq}
  V^{QQ} (r) \, = \, V_{adj}^\phi (r) + V_{adj}^{t^0} (r) - V_{adj}^B
  (r) + V_{adj}^G (r).
\end{align}
The right-hand-sides of Eqs.~\peq{netVqq} and \peq{netVr_qq} contain
terms given by the results of the above calculations of the
contributions to the $Q\bar Q$ potential.  The $QQ$ potential from
\eq{netVqq} can be evaluated numerically using the above techniques
and is plotted in \fig{totalpot} (upper line). It appears to be
attractive, though somewhat weaker than the $Q\bar Q$ potential.

The large-$r\, T$ behavior of $V^{QQ} (r)$ is dominated by the
graviton exchange, just like in the $Q\bar Q$ case: 
\begin{align}
  \label{netqq_large_r}
  V^{QQ} (r) \bigg|_{r \, T \gg 1} \, \approx \, V_{adj}^G (r)
  \bigg|_{r \, T \gg 1} \, \approx \, - \frac{\lambda}{2 \, N_c^2} \,
  \frac{0.86 \pm 0.03}{2 \, \pi \, r} \, e^{- 2.33 \, \pi \, r \, T}.
\end{align}

The small-$r\, T$ limit is less straightforward to find. First of all,
using Eqs.~\peq{Vdil-small-r}, \peq{Vt0-small-r}, \peq{KRpot_small-r}, and
\peq{GRpot_small-r} in \eq{netVr_qq} yields
\begin{align}
  \label{netqq-small-r}
  V^{QQ} (r) \bigg|_{r \, T \ll 1} \, = \, 0 \ \ \ \text{(in the
    leading--logarithmic approximation)}.
\end{align}
This result can be interpreted as follows. The large-$q$ asymptotics
of the supergravity fields given by Eqs.~\peq{dil_large-q},
\peq{t0_large-q}, \peq{KRsol_large-q}, and \eq{GRsol_large-q}, that we
used in deriving the potential asymptotics in Eqs.~\peq{Vdil-small-r},
\peq{Vt0-small-r}, \peq{KRpot_small-r}, and \peq{GRpot_small-r} in
\eq{netVr_qq} comes from the part of the EOM for those fields that is
independent of the curvature of AdS$_5$. Hence this part of the
string-string interactions is the same as in the flat space. In the
flat space, however, we know that the interaction between the parallel
strings oriented in the same direction should vanish since this would
be a BPS system. In a way the zero of \eq{netqq-small-r} provides a
cross-check that we have properly accounted for all the relevant
supergravity modes.

The zero on \eq{netqq-small-r} only implies that the
leading--logarithm terms containing $(1/r) \ln (1/r\, T)$, which gave
rise to the small-$r\, T$ potential \peq{net-small-r} in the $Q\bar Q$
case, cancel among each other in the $QQ$ case. This does not imply
that the $QQ$ potential actually goes to zero as $r \, T \rightarrow
0$. To determine the correct $r \, T \rightarrow 0$ asymptotics of the
$QQ$ potential we have to solve the EOM for the four relevant
supergravity fields ($\phi$, $t^0$, $B_{MN}$, $h_{MN}$) at $T=0$ and
integrate
\begin{align}
  \label{netVqq_T0}
  V^{QQ}_{T=0} (q) \, = \, \frac{\lambda}{2 \, N_c^2} \,
  \int\limits_0^\infty \, \frac{d z}{z^2} \, \left[ \varphi_{T=0} (q,
    z) + \frac{5}{3} \, \tgt^0_{T=0} (q, z) - 4 \, z^2 \, {\cal
      B}_{0z}^{T=0} (q, z) - \cA_{T=0} (q, z) \right].
\end{align}

Putting $z_h \rightarrow \infty$ in the dilaton EOM \peq{dil_EOM3} yields
\begin{align}
  \label{dil_EOM_T=0}
  \varphi_{z z} - \frac{3}{z} \, \varphi_z - q^2 \,
  \varphi = z,
\end{align}
with the solution
\begin{align}\label{dil_T=0}
  \varphi_{T=0} (q, z) \, = \, - z^3 \, \left\{ \frac{1}{3} +
    \frac{\pi}{2 \, q \, z} \, \left[ L_2 (q \, z) - I_2 (q \, z)
    \right] \right\}.
\end{align}
In arriving at \eq{dil_T=0} we have required that $\varphi (q, z=0)
=0$ and that $\varphi$ does not diverge more than linearly in $z$ as
$z \rightarrow \infty$ (see \eq{dil_large-q}). In \eq{dil_T=0}, $L_2$
is the modified Struve function and $I_2$ is the modified Bessel
function. Due to the linear divergence at large-$z$, the contribution
of $\varphi$ from \eq{dil_T=0} to the $QQ$ potential \peq{netVqq_T0}
is infinite: since we know that this infinity would cancel with
contributions from other fields, we have to add those other
contributions first before integrating the result over $z$.

For the massive scalar $t^0$ the $T=0$ EOM in momentum-space is
\begin{align}
  \label{t0_EOM_T0}
  \tgt^0_{z z} - \frac{3}{z} \ \tgt^0_z - q^2 \, \tgt^0 \,
  - \frac{32}{z^2} \, \tgt^0 \, = \, z.
\end{align}
Imposing the boundary condition $\tgt^0 (q, z=0) =0$ and requiring
that $\tgt^0$ is no more than linearly divergent in $z$ as $z
\rightarrow \infty$ (see \eq{t0_large-q}) we obtain the solution of
\eq{t0_EOM_T0}
\begin{align}
\label{t0T0}
\tgt^0_{T=0} (q, z) \, = \, - \frac{z^3}{35} \ F \left( 1;
  -\frac{3}{2}, \frac{9}{2} ; \frac{q^2 \, z^2}{4} \right) +
\frac{\pi}{2 \, q} \, z^2 \, I_6 (q \, z).
\end{align}
Here $F$ is the generalized hypergeometric function.

The zero-temperature EOM of motion for the 2-form field is
\begin{align}\label{KR_EOM_T0}
  z^2 \, {\cal B}''_{0z} + 3 \, z \, {\cal
    B}'_{0z} - (15 + q^2 \, z^2) \, {\cal B}_{0z} = z \ .
\end{align}
The series solution of this equation can be straightforwardly
constructed (again imposing the same boundary conditions as for the
scalars, with the exception that the 2-form field actually goes to
zero as $z \rightarrow \infty$, as follows from \eq{KRsol_large-q}).
The summation of the series is more involved, but can be accomplished
yielding
\begin{align}\label{BT0}
  q \, {\cal B}_{0z}^{T=0} (q,z) \, = \, - \frac{2 \, \pi \, i}{q \,
    z} \, I_4 (q \, z) + \frac{1}{4 \, q^2 \, z^2} \, \bigg\{ 2 \, \pi
  & \left[ ( 8 + q^2 \, z^2) \, I_1 (q \, z) - 4 \, q \, z \, \left( 1
      + 2 \, I_2 (q \, z) \right) \right] \, Y_4 (- i \, q \, z)
  \notag \\ & + \pi \, q^3 \, z^3 \ G^{3 \, 0}_{2 \, 4} \left( -
    \frac{i \, q \, z}{2} , \frac{1}{2} \bigg|
\begin{array}{cccc}
 & -\frac{5}{2}, & 0 & \\
 -2, & -1, & 2, & - \frac{5}{2}
\end{array}
\right) \bigg\}
\end{align} 
where $G$ is the generalized Meijer $G$-function. Despite several
terms containing an $i$ on the right-hand-side, \eq{BT0} gives a
real-valued 2-form field.

Finally, the Einstein equations at $T=0$, reduced to a single equation
for $\cA$, read
\begin{align}\label{GR_EOM_T0}
  z \, (3 - q^2 \, z^2) \, \cA_{z \, z} - 3 \, \left( 1 \, - q^2 \,
    z^2 \right) \, \cA_z - q^2 \, z \, \left[1 - q^2 \, z^2 \right] \,
  \cA \, = \, \frac{4}{3} \, q^2 z^4 \ .
\end{align}
This equation can be solved by series expansion similar to how it was
done in the non-zero temperature case. However, the series can only be
summed numerically. Instead we simply solve \eq{GR_EOM_T0}
numerically, requiring that $\cA (q, z=0) =0$ and that $\cA$ maps onto
the asymptotics of \eq{GRsol_large-q} at large-$q \, z$.

Numerical evaluation of the resulting $QQ$ potential at $T=0$ by the
integration over $z$ of the sum of all four contributions following
\eq{netVqq_T0} yields
\begin{align}
  \label{netVqq_T0_result}
  V^{QQ}_{T=0} (q) \, = \, (-0.7 \pm 0.1) \,
  \frac{\lambda}{N_c^2} \frac{1}{q^2}.
\end{align}
The corresponding coordinate-space potential is
\begin{align}\label{netVqq_T0_r}
  V^{QQ} (r) \bigg|_{r \, T \ll 1} \, = \, (-0.7 \pm 0.1) \,
  \frac{\lambda}{N_c^2} \frac{1}{4 \, \pi \, r}.
\end{align}
We see that the net $QQ$ potential at small-$r \, T$ is attractive. We
also observe that it is finite in the $T \rightarrow 0$ limit, unlike
the $Q\bar Q$ potential of \eq{net-small-r}, and does not require
string--fluctuation corrections to remain finite in this limit.

%%%%%%%%%%%%%%%%%%%%%%%%%%%%%%%%%%%%%%%%%%%%%%%%%%%%%%%%%%%%%%%%%%%%%%%%%%%%%%%

%%%%%%%%%%%%%%%%%%%%%%%%%%%%%%%%%%%%%%%%%%%%%%%%%%%%%%%%%%%%%%%%%%%%%%%%%%%%%%%

\section{Summary and outlook}
\label{sum}

We computed the adjoint quark--anti-quark potential, as well as the
quark--quark potential in a SYM plasma. Both potentials are of
order-$\lambda/N_c^2$, and both are attractive. For comparison, notice
that at weak coupling both potentials are also of
order-$\lambda/N_c^2$. Moreover, the adjoint $Q\bar Q$ potential is
repulsive at weak coupling, and so is the the $QQ$ potential with the
quarks in the $N_c^2 - N_c$ representation. On the other hand, the
$QQ$ potential in the $\bar{N}_c$ representation is attractive at weak
coupling. It appears that all the potentials, regardless of whether
they are repulsive or attractive, become attractive at large 't Hooft
coupling.

To interpret this result we use the suggestion of \cite{Arnold:1995bh}
that the electric modes can be singled out non-perturbatively by
identifying the $CT$-odd interaction channel. The quantum numbers of
the supergravity fields contributing to the potentials calculated
above can be found in \cite{Brower:2000rp} (see also \cite{Bak:2007fk}
for a table of the quantum numbers of the relevant fields excluding
$t^0$). We see that the dilaton, graviton, and $t^0$ are $CT$-even,
while $B_{0z}$ is $CT$-odd. We can thus identify the 2-form
contribution as corresponding to chromo-electric modes in the gauge
theory. The remaining dilaton, massive scalar and the graviton may
correspond to either magnetic modes, or to even numbers of electric
mode exchanges. The latter are probably not very important in the mix,
since the chromo-electric (2-form) contribution by itself is smaller
than the sum of all other contributions: hence it is likely that the
scalars and the dilaton correspond mainly to the chromo-magnetic modes
on the boundary. The chromo-electric (2-form) contribution changes
sign in going from $Q\bar Q$ to $QQ$, which seems natural for the
chromo-electric modes. In the perturbative weak-coupling limit, when
the potentials are calculated at lowest order by two-gluon exchanges
between Polyakov loops \cite{Nadkarni:1986cz}, the interaction is
entirely due to chromo-electric modes. However, even at weak coupling
it appears that the chromo-magnetic modes are less screened than the
chromo-electric ones, though they do not couple directly to the heavy
quark and anti-quark, and their contribution is suppressed by extra
powers of the coupling coming from the loop diagrams needed to couple
them to the quarks \cite{Braaten:1994qx,Arnold:1995bh}. It seems that
our above results suggest that at strong coupling the magnetic modes
dominate over electric modes in the $Q\bar Q$ and $QQ$ potentials,
making both of them attractive. The fact the the leading pole for the
graviton is smaller than the leading pole for the 2-form field can be
interpreted as magnetic modes being less screened even in the strong
coupling limit. Small-coupling suppression of the magnetic
contribution's coupling to the static heavy quarks by higher orders of
the coupling is no longer an issue at strong coupling. Thus it seems
natural that magnetic modes dominate at strong coupling (since they
are screened less than electric modes and couple to quarks similarly)
making both potentials attractive.

For the adjoint $Q\bar Q$ potential (and for the $QQ$ potential in the
$N_c^2 - N_c$ representation) we thus conjecture that in transition
from weak to strong coupling the potential, while remaining of
order-$\lambda/N_c^2$, would change its sign.  There should be no
change of sign for the $QQ$ potential in the $\bar{N}_c$
representation.

In \cite{Bak:2007fk} the Debye screening mass was identified as the
leading pole in the $CT$-odd channel. Of all the $CT$-odd supergravity
particles, the one with the lowest location of the pole on imaginary
axis is the axion \cite{Brower:2000rp}, the EOM and, therefore, poles
for which are the same as for the dilaton considered above. The
authors of \cite{Brower:2000rp} have therefore identified $m_D =
3.4041 \, \pi \, T$ as the Debye mass of ${\cal N} =4$ SYM plasma.
However, it may seem a little peculiar that the axion does not couple
to the string world-sheet at the leading (tree-level) order: how can
electric modes in gauge theory not couple to a static heavy quark? It
is possible (though seems a little hard to determine precisely) that
the single axion can still couple to the string through some
higher-order diagrams, with the axion-string coupling suppressed by
additional powers of $1/N_c^2$. If that was the case, the $CT$-odd
part of the $Q\bar Q$ potential at large-$rT$ would be proportional to
\begin{align}
  \frac{\lambda}{N_c^2} \, e^{- 5.1085 \, \pi \, r \, T} + O \left(
    \frac{1}{N_c^6} \right) \, e^{- 3.4041 \, \pi \, r \, T}.
\end{align}
Indeed at very large distances, parametrically defined by $r \, T \gg
\ln N_c^2$, the axion would dominate, with its mass being the correct
Debye mass. It appears that more work is needed in order to
determine the degree of suppression of a single-axion coupling to the
string world-sheet and to eliminate the possibility of this coupling
being zero. If the coupling is in fact zero, the leading pole of the 2-form field at $5.1085 \,
\pi \, T$ would be the correct Debye mass. We leave this for future work.

Other possible future improvements of our result may include
performing similar calculations of the $Q\bar Q$ and $QQ$ potential in
more QCD-like geometries
\cite{Polchinski:2001tt,BoschiFilho:2002vd,Erlich:2005qh,DaRold:2005zs,BoschiFilho:2005yh,Grigoryan:2007vg,Karch:2006pv,Karch:2010eg,Grigoryan:2007my,Andreev:2006nw,Andreev:2009zk,He:2010ye}.
Indeed the positions of the poles and the corresponding residues would
be modified by the new geometries, along with other calculational
details. One may even include finite quark masses (which are infinite
in our present calculation) by inserting a probe D7-brane wrapped over
$S^5$ along the lines of \cite{Karch:2002sh} and having the strings
dual to the quarks end on this D7 brane.  After performing a more
QCD-like calculations, one may hope to be able to perform a meaningful
comparison of the results to the lattice QCD data
\cite{Bazavov:2009us,Petreczky:2005bd}.

%%%%%%%%%%%%%%%%%%%%%%%%%%%%%%%%%%%%%%%%%%%%%%%%%%%%%%%%%%%%%%%%%%%%%%%%%%%%%%%

\acknowledgments

The authors are grateful to Dick Furnstahl, Samir Mathur, Jorge
Noronha, Misha Stephanov, Chung-I Tan, Diana Vaman, and Larry Yaffe for very
informative discussions. We would like to especially thank Dick
Furnstahl for the help with the numerical solution, and Samir Mathur
and Chung-I Tan for discussions of their work \cite{Brower:2000rp}.

This research is sponsored in part by the U.S. Department of Energy
under Grant No. DE-SC0004286.

%%%%%%%%%%%%%%%%%%%%%%%%%%%%%%%%%%%%%%%%%%%%%%%%%%%%%%%%%%%%%%%%%%%%%%%%%%%%%%%%%

\appendix

\renewcommand{\theequation}{A\arabic{equation}}
  \setcounter{equation}{0}
\section{Solution of Einstein equations with a straight string as a source}
\label{A}

Linearized Einstein equations \peq{ein} for the metric
\peq{metricgeneral} read
\begin{subequations}\label{Ein1}
  \begin{align}
    (t \, t) \hspace*{1cm} & 8 \, {\cal B} + z \, (1+z^4) \, (3 {\cal
      A}_z - {\cal B}_z) + 2 \, z \, (2+z^4) \, \tgh_z - z^2
    \, (1-z^4) \, \tgh_{z \, z} + q^2 \, z^2 \, \tgh
    \, = \, - \frac{2}{3} \, z^3 \label{Ein1a} \\
    (r \, r) \hspace*{1cm} & 8 \, \cB + z \left[ 2 \, (3-z^4) \, \cA_z
      - (1-z^4) \, \cB_z + (1-z^4) \, \tgh_z - z \, (1-z^4) \, \cA_{z
        \, z} + z \, q^2 \, \cA \right. \notag \\ & \hspace*{5cm}
    \left. - z \, r \, \partial_r^2 \, (\cA + \cB + \tgh) \right] \, =
    \, \frac{4}{3}
    \, z^3 \\
    (\theta \, \theta) \hspace*{1cm} & 8 \, \cB + z \left[ 2 \,
      (3-z^4) \, \cA_z - (1-z^4) \, \cB_z + (1-z^4) \, \tgh_z - z \,
      (1-z^4) \, \cA_{z \, z} + z \, q^2 \, \cA \right. \notag \\ &
    \hspace*{5cm} \left. - z \, \partial_r \, (\cA + \cB + \tgh)
    \right] \, = \, \frac{4}{3} \, z^3 \\
    (z \, z) \hspace*{1cm} & 8 \, \cB + z \left[ 3 \, (1+z^4) \, \cA_z
      - 2 \, (2-z^4) \, \cB_z + (1 + 5 \, z^4) \, \tgh_z - 3 \, z \,
      (1-z^4) \, \cA_{z \, z} - z \, (1-z^4) \, \tgh_{z \, z} \right.
    \notag \\ & \hspace*{5cm} \left. + \, z \, q^2 \, \cB \, \right]
    \, = \, - \frac{2}{3} \, z^3
  \end{align}
\end{subequations}
where we have omitted the $\phi \, \phi$ and $r \, z$ components,
which follow from the equations given above. As usual $\cA_z =
\partial_z \cA$ and $\cA_{z \, z} = \partial_z^2 \cA$.

Note that it appears difficult to perform Fourier transform on the $r
\, r$ and $\theta \, \theta$ components, since, for our $\theta$- and
$\phi$-independent metric in spherical coordinates not all
$r$-derivatives in those equations lead to powers of $q$ in an obvious
way. That is why we have, perhaps a bit sloppily, left some powers of
$r$ and $\partial_r$ in those two equations: these object should be
understood now as some operators in $q$-space. Subtracting $r \, r$
and $\theta \, \theta$ equations from each other, and transforming the
result back into coordinate space, we get
\begin{align}
  - r \, \partial_r^2 \, (\bar{A} + \bar{B} + \bar{h}) + \partial_r \,
  (\bar{A} + \bar{B} + \bar{h}) \, = \, 0 
\end{align}
which implies that
\begin{align}
  \bar{A} + \bar{B} + \bar{h} \, = \, \rho_1 (z) \, r^2 + \rho_2 (z)
\end{align}
with $\rho_1 (z)$ and $\rho_2 (z)$ some arbitrary functions of $z$. On
the physical grounds one may require that the metric should fall off
with increasing $r$ at large-$r$: this would put $\rho_1$ and $\rho_2$
to zero. Alternatively one can show that, employing the residual
diffeomorphism invariance left in the metric ansatz
\peq{metricgeneral} one can eliminate $\rho_1 (z)$ and $\rho_2 (z)$,
obtaining \eq{ABh}, which in momentum space leads to
\begin{align}
  \label{h_elim}
  \cA + \cB + \tgh \, = \, 0. 
\end{align}
Using \eq{h_elim} to eliminate $\tgh$ from Eqs.~\peq{Ein1}, and
subtracting $z \, z$ equation from $t \, t$, we write
\begin{subequations}\label{Ein2}
  \begin{align}
    (t \, t) - (z \, z) \hspace*{1cm} & - 3 \, (1-z^4) \, \cA_z + 3 \,
    z \, (1-z^4) \, \cA_{z \, z} - z \, q^2 \, \cA
    - 2 \, z \, q^2 \, \cB \, = \, 0  \label{Ein2a} \\
    (r \, r) \hspace*{1cm} & 8 \, \cB + z \left[ (5-z^4) \, \cA_z - 2
      \, (1-z^4) \, \cB_z - z \, (1-z^4) \, \cA_{z \, z} + z \, q^2 \,
      \cA \right] \, = \, \frac{4}{3} \, z^3, \label{Ein2b}
  \end{align}
\end{subequations}
which is now written entirely in momentum space. Solving \eq{Ein2a}
for $\cB$ yields
\begin{align}
  \label{BA}
  \cB \, = \, - \frac{1}{2 \, z \, q^2} \, \left[ 3 \, (1-z^4) \,
    (\cA_z - z \, \cA_{z \, z}) + z \, q^2 \, \cA \right]. 
\end{align}
Substituting \eq{BA} into \eq{Ein2b} yields a 3rd-order differential
equation for $\cA$. Solving this equation to express $\cA_{z\, z\, z}$
in terms of lower-order derivatives of $\cA$, we can use the result,
along with Eqs.~\peq{h_elim} and \peq{BA}, in, say, \eq{Ein1a} to
obtain \eq{GR_EOM} in the text (after some considerable algebra).

%%%%%%%%%%%%%%%%%%%%%%%%%%%%%%%%%%%%%%%%%%%%%%%%%%%%%%%%%%%%%%%%%%%%%%%%%%%%%%

%\bibliography{references} 
%\bibliographystyle{JHEP}

%%%%%%%%%%%%%%%%%%%%%%%%%%%%%%%%%%%%%%%%%%%%%%%%%%%%%%%%%%%%%%%%%%%%%%%%%%%%%%

\providecommand{\href}[2]{#2}\begingroup\raggedright\endgroup

%%%%%%%%%%%%%%%%%%%%%%%%%%%%%%%%%%%%%%%%%%%%%%%%%%%%%%%%%%%%%%%%%%%%%%%%%%%%%%

\end{document}